\documentclass[11pt,graphicx,amsmath]{article}
\usepackage{amsmath}
\usepackage{graphicx}
\usepackage{bm}
\usepackage[dvips]{color}
\usepackage{amssymb}
\usepackage{amsfonts}
\usepackage{comment}
\usepackage{cite}
\usepackage{todonotes}

\usepackage{caption}
\usepackage{subcaption}

\def\be{\begin{equation}}
\def\ee{\end{equation}}
\def\ba{\begin{eqnarray}}
\def\ea{\end{eqnarray}}
\def\nn{\nonumber}

\def\bl#1\el{\begin{align}#1\end{align}}
\def\l{\left}

\def\r{\right}

\title{  Point-splitting  regularization of
          the  stress  tensor of a coupling scalar field
          in de Sitter space }

\author{\small    Xuan Ye  \thanks{yyyyy@mail.ustc.edu.cn}  ,
                Yang  Zhang\thanks{yzh@ustc.edu.cn}  ,
               Bo Wang  \thanks{ymwangbo@ustc.edu.cn}  \\
 \small  Department of  Astronomy,  Key Laboratory
               for Researches in Galaxies and Cosmology, \\
 \small    School of Astronomy and Space Sciences, \\
 \small  University of Science and Technology of China,  Hefei, Anhui, 230026,  China \\
 }

 \date{}

\def\be{\begin{equation}}
\def\ee{\end{equation}}
\def\ba{\begin{eqnarray}}
\def\ea{\end{eqnarray}}
\def\nn{\nonumber}
\def\bl#1\el{\begin{align}#1\end{align}}

\begin{document}

\maketitle

\begin{abstract}

\large

We perform the point-splitting regularization
on the vacuum stress tensor of a coupling scalar field in de Sitter space
under the guidance from the adiabatically regularized Green's function.
For the massive scalar field  with the minimal coupling  $\xi=0$,
the  2nd order point-splitting regularization
yields a finite vacuum stress tensor
with a positive,   constant energy density,
which can be identified as the cosmological constant
that drives  de Sitter inflation.
For the coupling $\xi\ne 0$, we find that,
even if the regularized Green's function is continuous,  UV and IR convergent,
the point-splitting regularization does not
automatically lead to an appropriate  stress tensor.
The coupling $\xi R$ causes  log divergent terms,
as well as higher-order finite terms
which depend upon the path of the coincidence limit.
After removing these unwanted terms by extra treatments,
the 2nd-order  regularization
for small couplings $\xi \in(0,\frac{1}{7.04})$,
and  respectively the 0th-order regularization
for the conformal coupling $\xi=\frac16$,
yield a finite,  constant vacuum stress tensor, in analogy to the case  $\xi=0$.
For the massless   field with $\xi=0$ or $\xi=\frac16$,
the point-splitting regularization
yields a vanishing vacuum stress tensor,
and there is no conformal trace anomaly for $\xi=\frac16$.
If the 4th-order regularization were taken,
the regularized energy density for general $\xi$ would be negative,
which is inconsistent with the de Sitter inflation,
and the regularized Green's function would be singular at the zero mass,
which is unphysical.
In all these cases,
the stress tensor from the point-splitting  regularization
 is equal to that from the adiabatic one.
We also discuss the issue of
the adequate order of regularization.

\end{abstract}

\

PACS numbers:     98.80.Cq ,        04.62.+v ,    98.80.Jk , 95.30.Sf

    Inflationary universe, 98.80.Cq ;

    Quantum fields in curved spacetimes 04.62.+v ;

    Mathematical and relativistic  aspects of cosmology, 98.80.Jk;

    Relativity and gravitation,  95.30.Sf

\large

\section{Introduction}

The stress tensor and the Green's function of a quantum field in vacuum state
have  ultra-violet (UV) divergences due to the zero-point fluctuations.
Unlike in the Minkowski spacetime,
these divergences may not simply
be dropped \cite{UtiyamaDeWitt1962,FeynmanHibbs1965}
since the finite part of fluctuations has gravitational effects.
For instance, the vacuum energies of the inflationary scalar field
is a natural candidate for the driving source of inflation expansion,
and the vacuum fluctuations of the scalar field,
together with the perturbed metric field,
 can induce
the CMB anisotropies and polarization 
\cite{MaBertschinger1995,ZhaoZhang2006,XiaZhang2008,XiaZhang2009,YZhang2011,CaiZhang2012}.
To remove the UV divergences,
two classes of regularization methods have been proposed,
the point-splitting regularization
\cite{DeWitt1975,Christensen1976,AdlerLiebermanNg1977,
Christensen1978,BunchDavies1977,
BunchDavies1978,BunchChristensenFulling1978,Wald1978}
in the $x$-space,
and the adiabatic regularization in the momentum $k$-space
\cite{ZeldovichStarobinsky1972,ParkerFulling1974,FullingParkerHu1974,
HuParker1978,BLHu1978,Bunch1978,Birrell1978,Bunch1980,
AndersonParker1987,BunchParker1979,
BirrellDavies1982,ParkerToms,Parker2007,MarkkanenTranberg2013,Markkanen2018,
WangZhangChen2016,ZhangWangJCAP2018}.
The dimensional regularization
\cite{CandelasRaine1975,DowkerCritchley1976,DowkerCritchley1977,
BrownCassidy1977,Brown1977,Bunch1979},
and the zeta function  regularization
\cite{DowkerCritchley1976,DowkerCritchley1977,Hawking1977}
work in the $x$-space and
can be classified into the point-splitting.

The essence of a regularization program is to choose
an appropriate subtraction term
so that the regularized vacuum stress tensor be UV and IR convergent,
and respect the covariant conservation,
and,  for a massive scalar field,
the regularized vacuum energy density be positive.
In de Sitter space,  the regularized vacuum stress tensor
should also possess the maximum symmetry of de Sitter space.
The regularized spectral energy density and power spectrum
should also  be UV and IR convergent, and positive.
We refer to these as the desired properties of
a regularized vacuum stress tensor.
The adiabatic regularization
\cite{ZeldovichStarobinsky1972,ParkerFulling1974,FullingParkerHu1974,
HuParker1978,BLHu1978,Bunch1978,Birrell1978,Bunch1980,
AndersonParker1987,BunchParker1979,
BirrellDavies1982,ParkerToms,Parker2007,MarkkanenTranberg2013,Markkanen2018,
WangZhangChen2016,ZhangWangJCAP2018}
deals with  UV divergences in terms of the $k$-modes of a quantum field.
The subtraction term is systematically prescribed by the WKB approximate modes
to certain adiabatic order.
There is no universal recipe of regularization
for a general coupling.
Related to this,  Ref.\cite{ParkerFulling1974}
assumed the minimal subtraction rule
that only the minimum number of terms should be subtracted
to yield the convergent stress tensor.
In  the conventional prescription,
for a scalar field,
the 4th-order subtraction is used for the stress tensor \cite{ParkerFulling1974}
and the 2nd-order for the power spectrum \cite{Parker2007}.
However,  in Ref. \cite{ZhangYeWang2020}
we found that, for a massive scalar field in de Sitter space,
the 4th-order regularization leads to
a negative spectral energy density,
and that the conformal trace anomaly for $\xi=\frac16$  is an artifact
caused by the improper 4th-order subtraction term.
We also showed that,
the 2nd-order adiabatic regularization for  $\xi=0$
yields the positive, UV convergent spectral energy density and  power spectrum,
and so does the 0th-order regularization for  $\xi=\frac16$,
and there is no conformal trace anomaly
\cite{DowkerCritchley1976,Christensen1976,Hawking1977,
Christensen1978,Brown1977,DowkerCritchley1977,BrownCassidy1977,
BunchChristensenFulling1978,Wald1978}.
In Ref.\cite{ZhangWangYe2020} we have studied
the adiabatic regularization of a massless scalar field,
and the resulting stress tensor is zero for $\xi=0$ and $\xi=\frac16$,
agreeing  with the massless limit of the massive field \cite{ZhangYeWang2020}.
For these cases, the regularized spectral stress tensor
has been obtained  analytically,
from which follows
the regularized stress tensor by the numerical $k-$integratin.

The point-splitting regularization deals with the UV divergences in the $x$-space.
In this method,  the stress tensor is constructed from the Green's function
via differentiations and the coincidence limit.
If  the regularized Green's function is available,
one can use it to calculate the regularized stress tensor.
In literature, in lack of the full expression of regularized Green's function,
the unregularized Green's function is often expanded at small separation,
and several UV divergent terms are removed.
But this kind of naive subtraction would cause new IR divergences.
In Ref. \cite {ZhangYeWang2020},
for the scalar field with $\xi=0$ and $\xi=\frac16$ respectively,
we have obtained the analytical expression of
the regularized Green's function valid on the whole range of spacetime.
This has been achieved
by the  Fourier transformation of the adiabatically regularized power spectrum
of a pertinent order.

In this paper,  we shall perform the point-splitting regularization
on the  stress tensor,
using the 2nd-order regularized Green's functions
for the minimal-coupling $\xi=0$ and the small coupling $\xi >0$,
and  respectively  the 0th-order regularized Green's functions
for the conformal coupling $\xi=\frac16$ \cite{ZhangYeWang2020}.
In these cases,
we shall perform calculation in two equivalent schemes:
One scheme is to calculate the regularized stress tensor from
the regularized Green's function,
another is to calculate the unregularized,
and subtraction stress tensors
and then to take  their difference.
As we shall see,
given a well-defined, regularized Green's function with $\xi\ne 0$
may not automatically lead to an appropriate stress tensor,
and extra treatments are needed in both schemes.
Besides, we shall also perform
the 4th-order regularization for a general $\xi$
and point out the difficulties of its outcome.

In Sect. 2, we list the exact solution,
the power spectrum, and the Green's function
of the coupling  massive scalar field in the vacuum state in de Sitter inflation.

In Sect. 3, we list the adiabatically regularized Green's functions
to be used in Sections  4, 5, 6.

In Sect. 4,  for the  minimally-coupling $\xi=0$,
we use the 2nd-order adiabatically regularized Green's function
to calculate the vacuum stress tensor by the point-splitting method.
The regularized stress tensor is obtained
with a positive energy density.

In Sect. 5,  for a general coupling $\xi>0$,
we also adopt the 2nd-order point-splitting regularization.
By extra treatments,  we obtain the regularized stress tensor
which has a positive energy density
for small couplings $0 \leq \xi< \frac{1}{7.04}$
at a fixed parameter ratio $\frac{m^2}{H^2}=0.1$.

In Sect. 6, for the conformally-coupling $\xi = \frac{1}{6}$,
we adopt the 0th-order point-splitting regularization.
By the treatments analogous to Sect. 5,
we obtain the stress tensor with a positive energy density.

In Sect. 7,  as an examination,
we perform the 4th-order point-splitting regularization for a general $\xi$.
The regularized energy density is negative,
and the regularized Green's function is singular at the zero mass $m=0$.

Sect. 8  gives  the conclusions and discussions.

Appendix A  lists some formulae of differentiation which are used in the context.

Appendix B shows certain terms  which depend on the path of the coincidence limit.

\section{ The scalar field during de Sitter inflation}

The metric of  a flat Robertson-Walker spacetime  is
\be \label{metric}
ds^2=a^2(\tau)[d\tau^2- \delta_{ij}   dx^idx^j],
\ee
with  the conformal time $\tau$.
The Lagrangian density of  a massive scalar field $\phi$   is
\be
{\cal L} =\frac12 \sqrt{-g}(g^{\mu\nu}\phi_{,\mu}\phi_{,\nu}-m^2\phi^2-\xi R\phi^2) ,
\ee
and the  field  equation is
\be    \label{fieldequxi}
(  \Box +m^2 + \xi R   )\phi =0 ,
\ee
where $R= 6 a''/a^{3}$ is  the scalar curvature,
$m$ is the mass, and $\xi $ is a coupling constant.
For specific, we consider  $0 \leq \xi \leq \frac16$ in this paper.
The energy momentum tensor of the scalar field
is given by \cite{Bunch1980,AndersonParker1987}
\ba\label{Tmunu}
T_{\mu\nu} &  = &  (1-2\xi) \partial_ \mu \phi \partial_ \nu \phi
  +(2\xi - \frac12) g_{\mu\nu } \partial^\sigma  \phi \partial_\sigma  \phi
  -2\xi \phi_{;\mu\nu} \phi
  \nn \\
&&  + \frac12 \xi g_{\mu\nu} \phi \Box \phi
 -\xi (R_{\mu\nu}-\frac12 g_{\mu\nu} R + \frac32 \xi R g_{\mu\nu}) \phi^2
+ (\frac12 -\frac32 \xi )g_{\mu\nu }m^2 \phi^2  ,
\ea
satisfying  the conservation law $T^{\mu\nu}_{~~ ; \, \nu} = 0$.
Using the field equation \eqref{fieldequxi},
it can be also written as
\ba \label{Tmunue}
T_{\mu\nu}
&  = &  (1-2\xi) \partial_ \mu \phi \partial_ \nu \phi
  +(2\xi - \frac12) g_{\mu\nu } \partial^\sigma  \phi \partial_\sigma  \phi
  -2\xi \phi_{;\mu\nu} \phi
  \nn \\
&&  +  2 \xi g_{\mu\nu} \phi \Box \phi
 -\xi G_{\mu\nu} \phi^2
+ \frac12   g_{\mu\nu }m^2 \phi^2 ,
\label{tmnuexp}
\ea
with $G_{\mu\nu}=R_{\mu\nu}-\frac12 g_{\mu\nu} R$.
In the de Sitter space,  $G_{\mu\nu}=-\frac14 g_{\mu\nu}R$
and $R=12H^2$.
The trace of \eqref{tmnuexp} is
\bl
T^\mu\, _\mu
 = (6\xi -1) \partial^ \mu \phi \partial_ \mu \phi
          +\xi (1-6\xi) R \phi^2
          +2 (1-3\xi) m^2 \phi^2  ,
 \label{traceTmunu}
\el
in particular,
\bl \label{traceTmunu0}
T^\mu\, _\mu & =  - \partial^ \mu \phi \partial_ \mu \phi
              +2   m^2 \phi^2    ~~~~  \text{for $\xi=0$,}
\\
\label{traceTmunu16}
T^\mu\, _\mu & = m^2 \phi^2   ~~~~ \text{  for  $\xi=\frac{1}{6}$},
\el
where the equation  \eqref{fieldequxi} has been used.

The field operator can be written in terms of its Fourier modes
as the following
\be
\phi  ({\bf r},\tau) = \int\frac{d^3k}{(2\pi)^{3/2}}
        \left[ a _{\bf k}   \phi_k(\tau)  e^{i\bf{k}\cdot\bf{r}}
    +a^{\dagger}_{\bf k} \phi^{*}_k(\tau) e^{-i\bf{k}\cdot\bf{r}}\right]
\ee
where  $ a_{\bf k}, a^{\dagger}_{\bf k'} $ are
the annihilation and creation operators
and satisfy the canonical commutation relations.
Since the field equation is linear,
the $k$-modes are independent of each other.
In this paper we consider de Sitter space with a scale factor
\cite{Zhang1994,ZhangErXiaMiao2006}
\be \label{inflation}
a(\tau)=\frac{1}{H |\tau| },\,\,\,\,-\infty<\tau\leq \tau_1,
\ee
where $H$ is a constant, and
$\tau_1$ is the ending time of inflation,
and  the scalar curvature   $R  = 12 H^2$.
Let $\phi_k(\tau) =  v_k(\tau)/a(\tau)$.
Then the   equation of $k$-mode $v_k$   is
\be\label{equxi}
v_k'' + \Big[ k^2 +  \big( \frac{m^2}{H^2}
       +   12 \xi -2 \big) \tau^{-2} \Big] v_k = 0 .
\ee
The analytical  solution is
\be  \label{u}
v_k (\tau )  \equiv  \sqrt{\frac{\pi}{2}}\sqrt{\frac{x}{2k}}
  e^{i \frac{\pi}{2}(\nu+ \frac12) } H^{(1)}_{\nu} ( x) ,
\ee
with $ H^{(1)}_{\nu}$ being the Hankel function,
and the  conjugate $v_k^* $ is another independent solution,
where the variable   $x \equiv  k |\tau|$,  and
\be
\nu \equiv  \big(\frac94  - \frac{m^2}{H^2} -12 \xi \big)^{1/2} .
\ee
In  this paper we consider  $\nu$ being real.
In the high $k$ limit, the solution \eqref{equxi}
approaches the positive-frequency mode
$v_{k }(\tau ) \rightarrow \frac{1}{\sqrt{2k} }e^{-i  k\tau }$.
The Bunch-Davies (BD) vacuum  state
is defined as the state vector $|0 \rangle$ such that
\be\label{ask}
a_{\bf k}  |0 \rangle =0, ~~~
      {\rm for\   all} \  {\bf k} .
\ee
The  unregularized  Green's function in the  BD vacuum state is
given by the following
  \cite{CandelasRaine1975,DowkerCritchley1976,DowkerCritchley1977,
  BunchDavies1978,ZhangYeWang2020}
\ba
G(x^\mu - x'\, ^\mu)  & = &  \langle0|  \phi(x^\mu) \phi (x'\, ^\mu) |0\rangle
 =  \frac{1}{(2\pi)^3} \int d^3k \, e^{i \bf k \cdot (r-r')}
           \phi_k(\tau) \phi_{k'}^* (\tau ')  \nn \\
& = & \frac{|\tau|^{1/2} |\tau'|^{1/2} }{8 \pi  a(\tau)a(\tau')}
  \frac{1}{|r-r'|} \int_0^\infty  dk  k  \sin( k|r-r'|)
  H^{(1)}_{\nu} (k\tau ) H^{(2)}_{\nu} (k\tau') \nn \\
& = & \frac{H^2}{16 \pi^2}
\Gamma \big( \frac{3}{2}-\nu \big) \Gamma \big( \nu +\frac{3}{2} \big)
 \, _2 F_1  \left[\frac{3}{2}+\nu ,\frac{3}{2}-\nu ,2, ~  1 + \frac{\sigma}{2}
    \right] \, , \label{GreeHyper}
\ea
where $_2 F_1$ is the hypergeometric function \cite{Watson1958},
and
\be\label{sigmadef}
\sigma \equiv  \frac{1}{(2 \tau \tau')} \l[(\tau-\tau')^2-|{\bf r-r'|}^2 \r] \, .
\ee
The Green's function  satisfies the equation
\bl\label{eqGreen}
(\nabla_{\mu}\nabla^{\mu}+\xi R +m^2)G (x^\mu - x'\, ^\mu) = 0 .
\el
where  $\nabla_{\mu}$ is the covariant differentiation with respect to $x$.
It should be  remarked that
the analytic  expression \eqref{GreeHyper} is defined for $\nu\ne \frac32$
since the factor $\Gamma \big( \frac{3}{2}-\nu \big)$
is divergent at $\nu=\frac32$  ($m=0=\xi$),
for which the Green's function for $\nu=\frac32$ is given by \eqref{49}
 \cite{ZhangYeWang2020,ZhangWangYe2020}.
The Green's function at the equal time $\tau=\tau'$ is
\ba\label{Greenunregdf}
G({\bf r}- {\bf r}')
   =  \langle0|  \phi(\textbf{r},\tau) \phi (\textbf{r}',\tau) |0\rangle
   = \frac{1}{\bf |r-r'|} \int_0^\infty  \frac{\sin( k \bf |r-r'|)}{k^2}
          \Delta^2_k (\tau)  \, d k ,
\ea
and the auto-correlation function is
\ba\label{vevcorr}
G(0)= \langle0|  \phi(\textbf{r},\tau) \phi (\textbf{r},\tau) |0\rangle
         =   \frac{1}{(2\pi)^3} \int d^3k \,  |\phi_k(\tau)|^2
         = \int_0^{\infty} \Delta_k^2 (\tau)\frac{dk}{k}  ,
\ea
where the   power spectrum  is  the following
\ba \label{BunchDaviesSpectrum}
\Delta_k^2 (\tau)
& \equiv & \frac{ k^{3}}{2  \pi^2 a^2 }   |v_k(\tau)|^2
   =  \frac{H^2 }{8 \pi  }     x^3
      \big|  H^{(1)}_{\nu} ( x)   \big|^2 ,
\ea
which is nonnegative by definition.
At low $k$,  $\Delta_k^2   \propto k^{3-2 \nu}$,
giving an IR convergent  auto-correlation (\ref{vevcorr}) for $\nu< \frac32 $.
At  high $k$,
$\Delta_k^2   \propto
   k^{3} \big( \frac{1}{2k}  +\frac{4\nu^2 -1}{16  k^3 \tau^2}  \big)$,
leading  to  quadratic and  logarithmic  UV divergences of $G(0)$.
The corresponding  Green's function behaves as
$G(\sigma) \propto  \sigma ^{\nu -3/2}$ at large $\sigma$,
which is IR convergent for $\nu < \frac32 $.
$G(\sigma)$  is  UV divergent at $\sigma=0$.
In this paper we shall remove these UV divergences of the scalar field
by regularization.
(UV divergences also occur in the 2-point correlation function
of relic gravitational wave
\cite{WangZhangChen2016,ZhangWangJCAP2018},
 and  as well as in  the 2-point correlation function of
density perturbations 
\cite{Zhang2007,ZhangMiao2009,ZhangChenWu2019,ZhangLi2021,ZhangWu2022} 
in Gaussian approximation,
and we will not discuss these here.)

\section{ Adiabatic regularization of the Green's function }

There are two possible ways to remove the UV divergences of the Green's function.
One way is to adiabatically regularize the power spectrum
and the perform the Fourier transformation of the regularized power spectrum,
yielding the  adiabatically regularized  Green's function
which is both UV and IR convergent.
This has been done in Refs. \cite{ZhangYeWang2020,ZhangWangYe2020}.
Another way is to expand the Green's function at small distance
and to directly remove the UV divergences terms.
However, as pointed  in Refs. \cite{ZhangYeWang2020,ZhangWangYe2020},
this kind of regularization in the position-space does not work easily.
We give a brief  summary as the following.

The expansion of $G(\sigma)$
at small $\sigma $ generally has the following form
\[
G(\sigma) \simeq    \frac{H^2}{16 \pi^2}
 \Big[- \frac{2}{\sigma} +    (\frac{1}{4}-\nu ^2) \ln  \sigma
       + const    \Big]  +  O(\sigma)   ,
\]
where  $1/\sigma$ and   $\ln  \sigma$ are UV divergent and should be removed.
Often the exact subtraction terms for the Green's function
are not known,
so that a Hadamard type of function is usually assumed
  as the following
\cite{CandelasRaine1975,DowkerCritchley1976,
Brown1977,BrownCassidy1977,DowkerCritchley1977,Christensen1976,Christensen1978,BunchDavies1978,Bunch1979}
\be\label{subGreens}
G(\sigma)_{sub} = \frac{H^2}{16 \pi^2}
\Big[  - \frac{2}{\sigma} +  (\frac{1}{4}-\nu ^2) \ln  \sigma   \Big] ,
\ee
as an approximation to the exact subtraction terms at small distance.
Some constant term can be added to \eqref{subGreens}
in order to ensure the covariant conservation.
Although the $\ln \sigma$ subtraction term of  \eqref{subGreens}
removes the log UV divergence at $\sigma=0$, nevertheless,
it will also cause a new IR divergence at $\sigma=\infty$
in the regularized Green's function
\be\label{Grregx}
G(\sigma)_{reg} = G(\sigma)  -G(\sigma)_{sub}
\ee
 for general $m$ and $\xi$.
That is, the conventional Hadamard function \eqref{subGreens}
as a subtraction term
is not valid at large  $\sigma$,
and one still lacks exact subtraction terms
defined on the whole range of $\sigma$.
In general, such an exact subtraction term
is hard to find directly in  position space.
(However, for the special case $m=\xi=0$,
the exact subtraction term is  easy to find,
and the new $\ln \sigma$ IR divergence will not occur.
See the paragraph around  \eqref{49} later.)
To find such an exact subtraction term for a general $\xi$,
one can be assisted by the adiabatically  regularized power spectrum
defined in $k$-space. 
As shown in Refs.\cite{ZhangYeWang2020,ZhangWangYe2020},
for the cases $\xi=0$ and $\xi= \frac16$,
subtracting off the UV divergent terms of the power spectrum in $k$-space
to an appropriate adiabatic order
gives a regularized, UV and IR convergent power spectrum.
Then by the Fourier transformation of the regularized power spectrum,
one obtains the adiabatically regularized,
UV and IR convergent Green's function.
In this approach,
the appropriate subtraction term for the Green's function
is the Fourier transformation of the subtraction term
to the power spectrum,
and is valid on the whole range of $\sigma$.
As it turns out,
its functional form is not simple,
and quite different from the  Hadamard function  \eqref{subGreens}.
In this paper,
such kind of adiabatically regularized Green's functions
will be used in the point-splitting regularization,
and we list the relevant formulae of  the 2nd-order for $\xi=0$
and the 0th-order for $\xi=\frac16$ in the following.

\begin{figure}[htb]
\centering
\subcaptionbox{}
    {  \includegraphics[width = .45\linewidth]{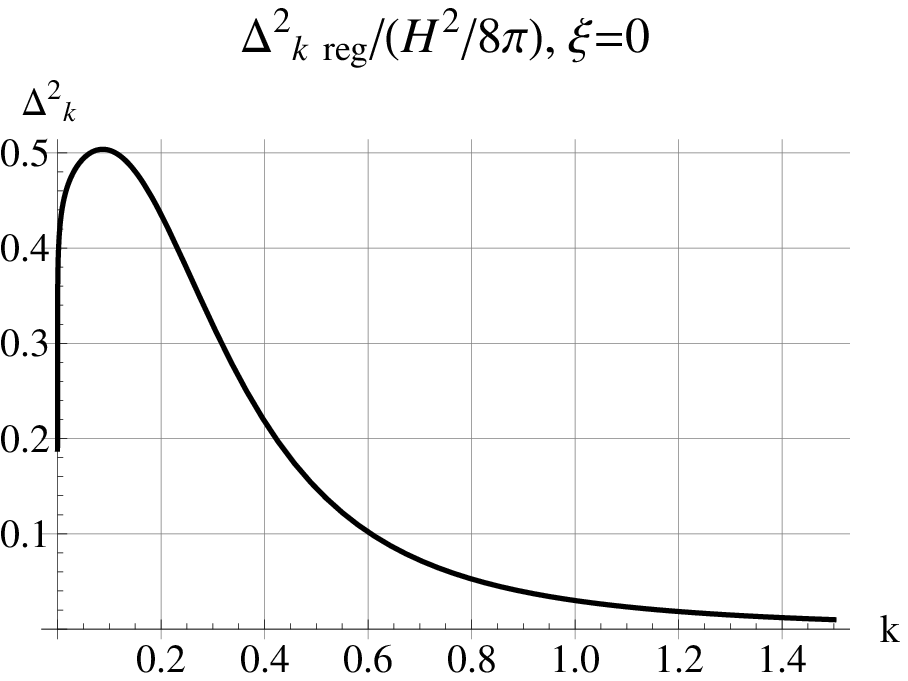}   }
\subcaptionbox{}
    {  \includegraphics[width = .5\linewidth]{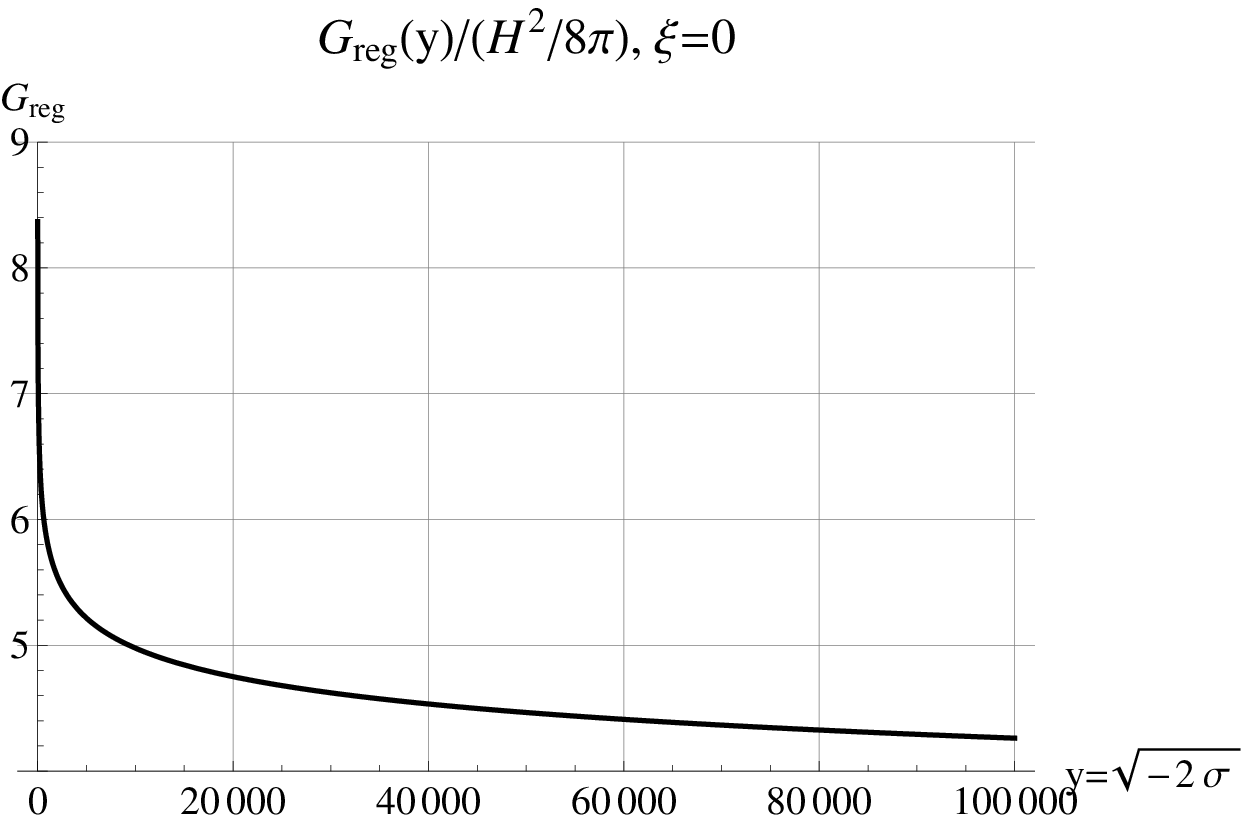}    }
\caption{ (a):
The 2nd-order regularized $\Delta^2_{k\, reg}$
in \eqref{PSxi162} is IR and UV convergent.
          (b): The 2nd-order regularized $G(y)_{reg }$
in (\ref{Greenreg2})  is  IR and UV convergent.
The model  for $\xi=0$ and $\frac{m^2}{H^2}=0.1$.
Here  $y=|r-r'|$ and $|\tau|=|\tau'|=1$ for illustration.
}
 \label{PSreg2nd}
\end{figure}

The adequate regularization depends upon the coupling $\xi$.
We assume the minimal subtraction rule
that only the minimum number of terms should be subtracted
to yield the convergent power spectrum.
This was originally assumed for
the stress tensor in Ref.\cite{ParkerFulling1974}.
For the minimally coupling $\xi=0$,
the 2nd-order regularization is adopted,
and the regularized power spectrum   is
\ba \label{PSxi162}
\Delta^2_{k\, reg} & = & \frac{ k^{3}}{2  \pi^2 a^2 }
      \Big( |v_k |^2 -|v_k^{(2)} |^2 \Big)
    =  \frac{ k^{3}}{2  \pi^2 a^2 }
      \Big( |v_k |^2 -\frac{1}{2 W^{(2)}} \Big) ,
\ea
where the 2nd-order effective inverse frequency
\be\label{W2xi}
\frac{1}{ W_k^{(2)}}
 =  \frac{1}{\omega}
       - 3  (\xi-\frac16)  \frac{a''/a}{\omega^{3}}
 + \frac{ m^2 (a'' a+ a'\,  ^2 )}{4 \omega^5}
      -  \frac{ 5 m^4 a'\, ^2 a^2 }{8\omega^7 }
\ee
with $\omega =\sqrt{k^2+m^2 a^2 }$.
(See (a29) in Ref.\cite{ZhangYeWang2020}.)
Replacing  $\Delta^2_{k}$ by $\Delta^2_{k\, reg}$
in eq.\eqref{Greenunregdf} yields
the 2nd-order regularized  Green's function for $\xi=0$ \cite{ZhangYeWang2020}
\ba\label{Greenreg2}
G(y)_{reg}
   & = &  G(y)  -G(y)_{sub} \nn \\
& = &  \frac{H^2}{8\pi} \frac{1}{y} \int_0^\infty d k
  \,  k \sin( k y) \Big( |H^{(1)}_\nu(k)|^2
  -\frac{2}{\pi}\frac{1}{ W^{(2)}} \Big) ,
\ea
where  $y=\bf |r-r'|$, $\nu =  (\frac94  - \frac{m^2}{H^2} )^{1/2}$,
  $G(y)$ is given by  eq.(\ref{GreeHyper}),
and  the subtraction   Green's function   is given by
\ba \label{sungrxi0}
G(y)_{sub}
& =  \frac{H^2}{4\pi^2} \Big[
        \frac{m}{H} \, \frac{1}{y} K_1 \Big(  \frac{m}{H} y \Big)
    + (1 - 6 \xi )   K_0 \Big(  \frac{m}{H} y \Big)  + \frac{1}{4}
    \frac{m}{H} \, y  K_1 \Big(  \frac{m}{H} y \Big)
     - \frac{1}{ 24 }  \frac{m^2}{H^2}
       y^2  K_2 \Big(  \frac{m}{H} y \Big) \Big] ,
\ea
with   $K_0$, $K_1$,  and $K_2$ being  the modified Hankel functions.
The   expression  \eqref{sungrxi0}
differs from the naive expression \eqref{subGreens}.
Fig.\ref{PSreg2nd} (a) shows that the 2nd-order regularized
$\Delta^2_{k\, reg}$ is  positive,  UV convergent and IR finite.
Fig.\ref{PSreg2nd} (b) shows that the resulting $G(y)_{reg}$,
is  UV finite and   IR convergent.

\begin{figure}[htb]
\centering
\subcaptionbox{}
    {  \includegraphics[width = .45\linewidth]{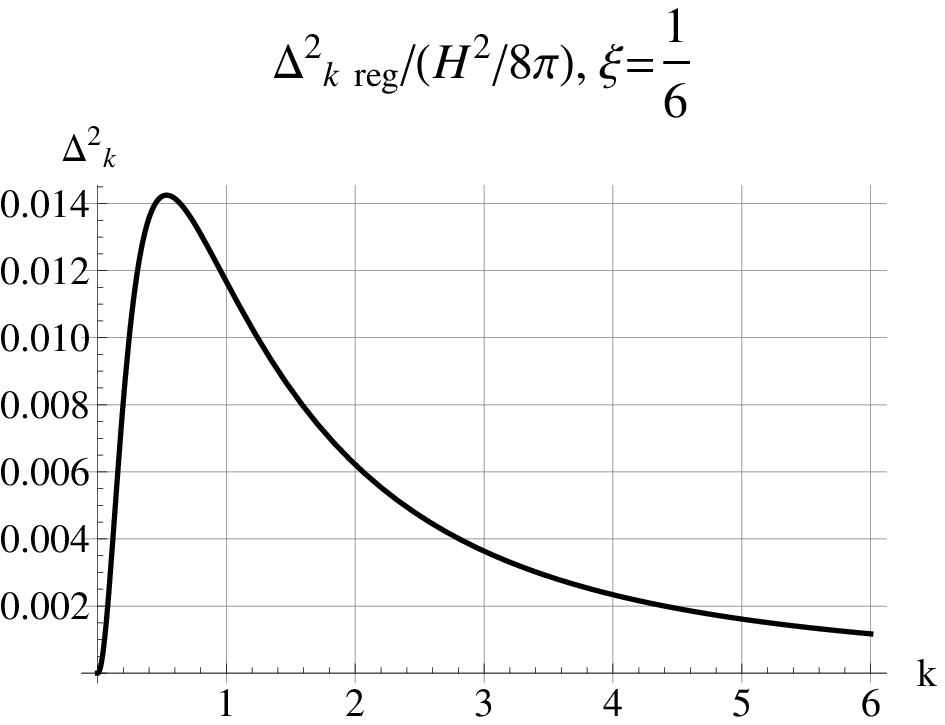}   }
\subcaptionbox{}
    {  \includegraphics[width = .48\linewidth]{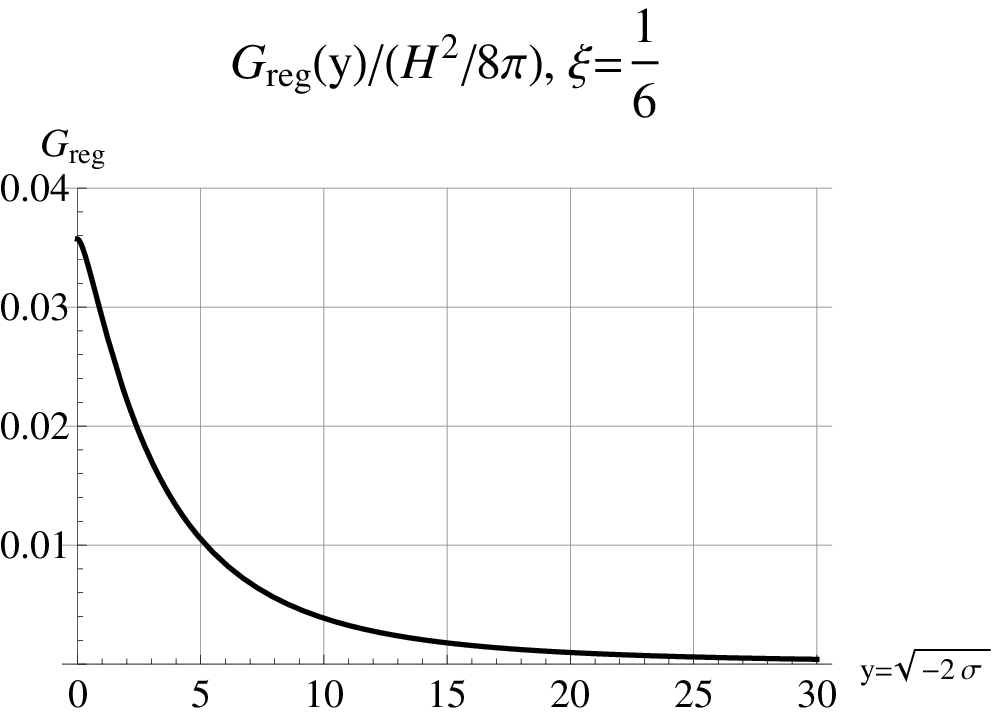}    }
\caption{ (a):  The 0th-order regularized $\Delta^2_{k\, reg}$
in \eqref{PSxi1600} is IR and UV convergent.
          (b):  The 0th-order regularized $G(y)_{reg }$
in (\ref{Greentrasfreg16})  is  IR and UV convergent.
The model  $\xi=\frac16$ and $\frac{m^2}{H^2}=0.1$.
}
 \label{PSreg0th}
\end{figure}

For the conformally coupling $\xi=\frac16$,
    the 0th-order regularization is adopted
\ba \label{PSxi1600}
\Delta^2_{k\, reg} & = & \frac{ k^{3}}{2  \pi^2 a^2 }
      \Big( |v_k |^2 -|v_k^{(0)} |^2 \Big)
    =  \frac{ k^{3}}{2  \pi^2 a^2 }
      \Big( |v_k |^2 -\frac{1}{2 W^{(0)}} \Big) ,
\ea
where $1/W^{(0)}=1/\omega $.
The 0th-order  regularized  Green's function for $\xi=\frac16$ is
\ba \label{Greentrasfreg16}
G(y)_{reg} & = &  G(y)  -G(y)_{sub}
\nn \\
& = & \frac{H^2}{8\pi} \frac{1}{y} \int_0^\infty d k
  \,  k \sin( k y) \Big( |H^{(1)}_\nu(k)|^2
   -\frac{2}{\pi}\frac{1}{\omega } \Big) ,
\ea
where  $\nu =   (\frac14  - \frac{m^2}{H^2} )^{1/2}$ for $\xi=\frac16$,
and the subtraction Green's function  is given by
\be\label{grsbxi16}
G(y)_{sub}  = \frac{H^2}{4\pi^2}
  \frac{m}{H} \, \frac{1}{y}   K_1\Big( \frac{m}{H} y \Big)  ,
\ee
also differing  from the naive expression \eqref{subGreens}.
Fig.\ref{PSreg0th} (a) shows that the 0th-order regularized
$\Delta^2_{k\, reg}$ is positive,  UV convergent and IR finite.
Fig. \ref{PSreg0th} (b) shows that $G(y)_{reg}$  is UV finite and IR convergent.

For the  nonequal time $\tau\ne \tau'$,
by the maximal symmetry in de Sitter space,
we just replace the variable $y \rightarrow  \sqrt{-2\sigma}$
in the expressions  \eqref{Greenreg2} \eqref{Greentrasfreg16}
and get the Green's functions $G_{reg}(x-x')$.
In the following sections
we shall use the regularized Green's functions
to calculate the regularized stress tensor
by the point-splitting method.

\section{The 2nd-order regularized  stress tensor with $\xi=0$}

\begin{figure}[htb]
\centering
\subcaptionbox{}
    {  \includegraphics[width = .45\linewidth]{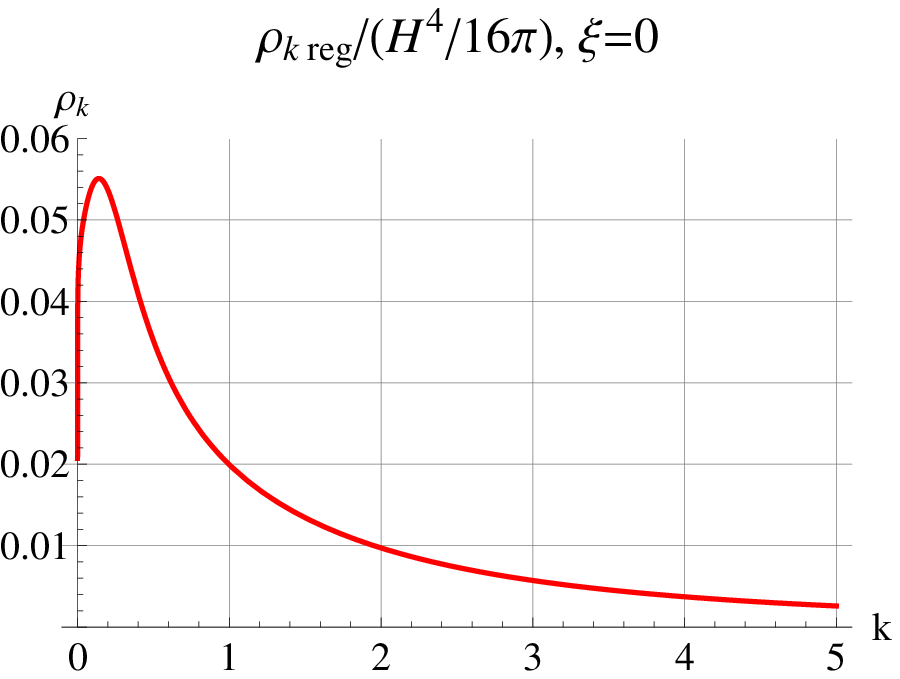}   }
\subcaptionbox{}
    {  \includegraphics[width = .48\linewidth]{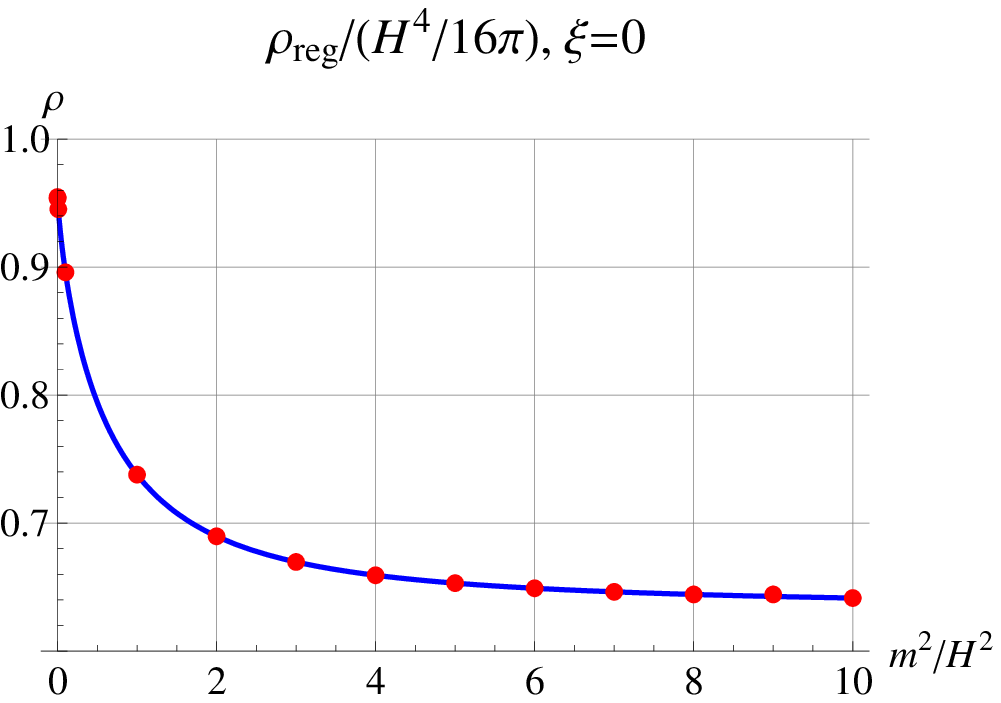}    }
\caption{ (a): The 2nd-order regularized
  spectral energy density $\rho_{k\, reg }$ in \eqref{energykxi0re}
  is positive, IR and UV convergent.
  The model $\xi=0$ and $\frac{m^2}{H^2}=0.1$.
   (b):  For $\xi=0$ the 2nd-order regularized energy density $\rho_{reg}$
   is positive and finite for the whole range of $\frac{m^2}{H^2}$.
   Blue line: the point-splitting \eqref{Lambda0};
   Red dots: the adiabatic  \eqref{intrhoreg}.
}
 \label{rho2ndregxi0}
\end{figure}

The stress tensor of the scalar field
contains UV divergences,
which must be subtracted by regularization.
Our goal is to achieve a regularized stress tensor
with the desired properties mentioned in the introduction.
In this section we study  the $\xi=0$ scalar field.
For a clear comparison,
we  first summarize briefly  the resulting stress tensor
from adiabatic regularization \cite{ZhangYeWang2020,ZhangWangYe2020},
and then present the point-splitting regularization in details,
using the regularized Green's function of the last section.
The energy density and pressure of the  $\xi=0$ scalar field
in the BD vacuum state
are  given by the expectation values
\be \label{energyspectr}
\rho  = \langle 0|  T^0\, _0  |0 \rangle =\int^{\infty}_0   \rho_k \frac{d k}{k} ,
\ee
\be
p =  -\frac13    \langle 0| T^i\, _i  |0 \rangle
             =  \int^\infty_0   p_k \frac{dk}{k}   ,
\ee
where the spectral energy density and the spectral pressure are
\ba \label{rhok}
\rho_k  &= & \frac{ k^3}{4\pi^2 a^4}
 \Big[ |v_k'|^2 + k^2  |v_k|^2 +m^2 a^2 |v_k|^2
  + (6\xi-1) \Big(  \frac{a'}{a} (v'_k v^*_k + v_k v^*\, '_k  )
    -  \frac{a'\, ^2}{a^2}  |v_k|^2  \Big) \Big] ,
   \\
p_k & = & \frac{k^3}{4 \pi^2 a^4}
  \Bigg[   \frac13 |v_k'|^2 + \frac13 k^2  |v_k|^2 - \frac13 m^2 a^2 |v_k|^2
  + 2(\xi-\frac16)\Big( -2 |v_k'|^2 + 3 \frac{a'}{a} (v'_k v^*_k + v_k v^*\, '_k  )
         \nn \\
&&  - 3(\frac{a'}{a})^2 |v_k|^2
    +  2(k^2 + m^2 a^2) |v_k|^2
    + 12 \xi  \frac{a''}{a} |v_k|^2  \Big)\Bigg]  .
     \label{sprectpressure}
\ea
$\rho_k$ is nonnegative,
and $p_k$  can take both positive and negative values.
For the minimal coupling  $\xi=0$ they reduce to
\ba\label{energykxi0}
\rho_k
& =& \frac{ k^3}{4\pi^2 a^4}
 \Big(  |(\frac{v_k}{a})' |^2  + k^2  |\frac{v_k}{a} |^2
 +m^2 a^2 |\frac{v_k}{a} |^2  \Big)   ,
         \\
p_k
&= & \frac{ k^3}{4\pi^2 a^4}
 \Big(   |(\frac{v_k}{a})' |^2  - \frac13 k^2 |\frac{v_k}{a}|^2
  - m^2 a^2 |\frac{v_k}{a}|^2  \Big)  . \label{pk}
\ea
At low $k$, $ \rho_k $ and $ p_{k}$
are IR convergent and dominated by the mass term.
At high $k$, $ \rho_k $ and $ p_{k}$
 contain  quartic, quadratic, and logarithmic UV divergences
which are removed by adiabatic regularization.
The 2nd-order adiabatic  regularization is performed
as the following \cite{ZhangYeWang2020}
{\allowdisplaybreaks
\ba\label{energykxi0re}
\rho_{k\, reg}
& = &   \rho_k - \rho_{k\,A 2} \nn \\
&=& \frac{ k^3}{4\pi^2 a^2}
 \Big(  |(\frac{v_k}{a})' |^2  + k^2  |\frac{v_k}{a} |^2
 +m^2 a^2 |\frac{v_k}{a} |^2  \Big) \nn \\
 & & - \frac{k^3}{4\pi^2 a^4}
\Big[  \omega + \frac{m^4 a^4}{8\omega^5} \frac{a'\,^2}{a^2}
    + (\xi -\frac16)  \big( -\frac{3}{\omega} \frac{a'\,^2}{a^2}
                    - \frac{3m^2 a'\, ^2}{\omega^3}  \big) \Big]  ,
\ea
\ba
p_{k\, reg}  & = &  p_k - p_{k\,A 2}
\nn \\
& = & \frac{k^3}{4 \pi^2 a^2}
  \Big(   |(\frac{v_k}{a})' |^2  - \frac13 k^2 |\frac{v_k}{a}|^2
  - m^2 a^2 |\frac{v_k}{a}|^2  \Big)  \nn \\
& &  -  \frac{k^3}{12 \pi^2 a^4}  \Big[
\omega - \frac{m^2 a^2}{\omega }
- \frac{m^4 a^4}{8\omega^5} (  \frac{2 a''}{a}+\frac{a'\,^2}{a^2})
  +\frac{5 m^6 a^6}{8 \omega^7} \frac{a'\,^2}{a^2} \nn \\
&&      + (\xi -\frac16)  \Big( \frac{1}{\omega} (6\frac{a''}{a}- 9\frac{a'\,^2}{a^2})
       +\frac{6m^2 a^2}{\omega^3}(\frac{a''}{a}-\frac{a'\,^2}{a^2})
      - \frac{9m^4 a^4}{\omega^5}\frac{a'\,^2}{a^2} \Big)
        \Big]   . \label{presskreg}
\ea
}
For the minimally-coupling,
we just set $\xi=0$ in \eqref{energykxi0re} and \eqref{presskreg}.
Both regularized spectra $\rho_{k\, \, reg}$ and $p_{k\, \, reg}$
are UV and IR convergent, respect the covariant conservation.
$\rho_{k\, \, reg}$ is positive,
and $p_{k\, reg}$  can take negative values.
Fig.\ref{rho2ndregxi0} (a) plots
$\rho_{k\, reg}$ for the model $\frac{m^2}{H^2}=0.1$.
(As demonstrated in Ref.\cite{ZhangYeWang2020},
for $\xi=0$,
the 0th-order regularization would not be able to remove all the UV divergences,
and
the 4th-order would remove too much, causing a spectral negative energy density.)

The 2nd-order adiabatically
regularized stress tensor  with  $\xi=0$ are obtained
 by the following $k$-integrations
\bl
 \rho_{reg} =\int_0^\infty ( \rho_{k} -\rho_{k\, A2})  \frac{dk}{k},
 ~~~~~
 p_{reg} = \int_0^\infty ( p_{k} -p_{k\, A 2} ) \frac{dk}{k} ,
 \label{intrhoreg}
\el
Interestingly, although the regularized spectra $\rho_{k\, reg} \ne  - p_{k\, reg}$,
nevertheless,  $\rho_{ reg} = - p_{ reg}$ after $k$-integration,
That is, the regularized stress tensor in the vacuum satisfies
the maximal symmetry in de Sitter space,
$\langle T_{\mu\nu } \rangle_{reg} = g_{\mu\nu}  \rho_{\,reg}$.
For instance,
$\rho_{reg} = - p_{reg}
 \simeq  0.895913 \frac{H^4}{16\pi} =89.5913 \frac{m^4}{16\pi} >0$
at $\frac{m^2}{H^2}=0.1$,
and  $\rho_{reg}      =- p_{reg}
\simeq  0.860342 \frac{H^4}{16\pi }=21.5086 \frac{m^4}{16\pi } $
at $\frac{m^2}{H^2}=0.2$.
We plot $\rho_{reg}$ as a function of
$m^2/H^2$ in red dots in Fig.\eqref{rho2ndregxi0} (b).

In the  massless limit $m=0$,
the adiabatically  regularized spectra  (\ref{energykxi0re}) and (\ref{presskreg})
become zero
\be \label{m=0conform}
\rho_{k\, \, reg}  =0 = p_{k\, \, reg}
~~~~~\text{for  $m=\xi=0$} .
\ee
and  the adiabatically  regularized stress tensor becomes \cite{ZhangYeWang2020,ZhangWangYe2020}
\be \label{m=0conformint}
\langle T_{\mu\nu}\rangle_{reg}=0,
~~~~~\text{for  $m=\xi=0$}.
\ee
It should be remarked that
the result \eqref{m=0conformint} comes from the ordering:
first the massless limit, then the $k$-integration.

In the point-splitting method
\cite{DeWitt1975,DowkerCritchley1976,Christensen1976,Christensen1978},
the vacuum stress tensor is calculated
by use of the Green's function in  $x$-space.
In the first  scheme one calculates
\bl\label{gnt}
\langle T_{\mu\nu}\rangle_{reg}
= & \lim_{x' \rightarrow x}\Big[ \frac12(1-2\xi)
   ( \nabla_{\mu}\nabla_{\nu'}+\nabla_{\mu'}\nabla_{\nu} )
    +(2\xi-\frac12)g_{\mu\nu}\nabla_{\sigma}\nabla^{\sigma'}
    -\xi ( \nabla_{\mu}\nabla_{\nu}+\nabla_{\mu'}\nabla_{\nu'} )
       \nn
\\
& +\xi g_{\mu\nu} (\nabla_{\sigma}\nabla^{\sigma}
  +\nabla_{\sigma'}\nabla^{\sigma'} ) -\xi G_{\mu\nu}
  +\frac12m^2g_{\mu\nu}\Big]   G_{reg}(x-x') \, ,
\el
where $G_{reg}(x-x') $  is the adiabatically regularized Green's function
given by  \eqref{Greenreg2} for $\xi=0$,
and is a biscalar at $x$ and at $x'$,
and $\nabla_\mu$ and  $\nabla_{\mu'}$
are the covariant differentiation
with respect to $x$ and $x'$ respectively.

Alternatively, in the second scheme,
one  calculates  the unregularized stress tensor
\bl\label{unrgrtmunmu}
\langle T_{\mu\nu}\rangle
 = &\lim_{x' \rightarrow x}\Big[ \frac12(1-2\xi)
     [\nabla_{\mu}\nabla_{\nu'}+\nabla_{\mu'}\nabla_{\nu}]
       +(2\xi-\frac12)g_{\mu\nu}\nabla_{\sigma}\nabla^{\sigma'}
       -\xi[\nabla_{\mu}\nabla_{\nu}+\nabla_{\mu'}\nabla_{\nu'}]
             \nn   \\
&+\xi g_{\mu\nu}[\nabla_{\sigma}\nabla^{\sigma}
+\nabla_{\sigma'}\nabla^{\sigma'}]-\xi G_{\mu\nu}
+\frac12m^2g_{\mu\nu}\Big]   G(x-x') \, ,
\el
and the subtraction stress tensor
\bl\label{subgrtmunmu}
\langle T_{\mu\nu}\rangle_{sub}
 = &\lim_{x' \rightarrow x }\Big[ \frac12(1-2\xi)
     [\nabla_{\mu}\nabla_{\nu'}+\nabla_{\mu'}\nabla_{\nu}]
       +(2\xi-\frac12)g_{\mu\nu}\nabla_{\sigma}\nabla^{\sigma'}
       -\xi[\nabla_{\mu}\nabla_{\nu}+\nabla_{\mu'}\nabla_{\nu'}]
             \nn   \\
&+\xi g_{\mu\nu}[\nabla_{\sigma}\nabla^{\sigma}
+\nabla_{\sigma'}\nabla^{\sigma'}]-\xi G_{\mu\nu}
+\frac12m^2g_{\mu\nu}\Big]   G(x-x')_{sub} \, ,
\el
with $G(x-x')_{sub}$  being the subtraction  Green's function,
and then takes the difference
\be
\langle T_{\mu\nu}\rangle_{reg}
 = \langle T_{\mu\nu}\rangle - \langle T_{\mu\nu}\rangle_{sub} \, .
\ee
The second scheme is often  adopted in literature
\cite{DeWitt1975,DowkerCritchley1976,Christensen1976,Christensen1978,BunchDavies1978}.
As we shall see, both schemes lead to the same result.

Now we calculate the stress tensor \eqref{gnt}
for $\xi=0$ by the first scheme of the point-splitting method.
For this purpose,  we only need
the 2nd-order regularized $G(\sigma)_{reg}$ at small separation
up to the order $\sigma^2$.
By the  maximal symmetry  in de Sitter space,
we replace   $y \rightarrow  \sqrt{-2\sigma}$
in \eqref{GreeHyper} and \eqref{sungrxi0},
and expand them at small separation
\bl
G(\sigma)
  =   \frac{1}{16 \pi^2}
\Big(-\frac{1}{\epsilon^2}+ W \ln\epsilon^2
   + X + Y \epsilon^2\ln\epsilon^2+ Z   \epsilon^2\Big)
   + O(\epsilon^3),    \label{grgenr}
\\
G(\sigma )_{sub}
=\frac{1}{16\pi ^2}\Big(-\frac{1}{\epsilon^2 }
  + W \ln\epsilon^2 + A + Y \epsilon^2\ln\epsilon^2
  +B \epsilon^2 \Big)
   + O(\epsilon^3)  ,  \label{subGxi0}
\el
where  $\epsilon^2 \equiv  \sigma/ 2H^2$  for simple notation,
and the constants are
\bl
W&=m^2+(\xi-\frac16)R ,
 \label{RrW276}
\\
X &=\big(m^2+(\xi-\frac16)R \big) \Big(-1+2\gamma + \ln(- H^2)
   +\psi(\frac{3}{2}-\nu)+\psi(\frac{3}{2}+\nu)\Big) ,
   \label{X}
\\
Y &  =-\frac12 \big(m^2+(\xi-\frac16)R \big) (m^2+\xi R) ,
\label{Y}
\\
Z & = -\frac12 \Big( m^2+(\xi-\frac16)R \Big)
(m^2+\xi R) \Big(-\frac52+2\gamma +\ln(- H^2)
   + \psi(\frac{5}{2}-\nu)+ \psi(\frac{5}{2}+\nu)\Big),
 \label{RrZ279}
 \\
 A & =  (m^2 -\frac{1}{6} R) \big(-1+2\gamma +\ln(- m^2) \big) -\frac{R}{9} \, ,
\nn \\
 B & = - \frac{1}{2} m^2 (m^2 - \frac{1}{6} R )
\big(-\frac52 +2 \gamma  +\ln(- m^2) \big)    -\frac{5 m^2 R}{72} \,  ,
\nn
\el
where  the psi function
$\psi(z) \equiv  \Gamma'(z)/\Gamma(z)$  with $z\ne 0, -1, -2$
\cite{GradshteynRyzhik1980,NISTHandbook2010}.
Both expressions  \eqref{grgenr} \eqref{subGxi0} have a similar structure,
and their difference is
the 2nd-order adiabatically  regularized Green's function for $\xi=0$
at small separation
\bl
G(\sigma )_{reg}
 =&   \frac{1}{16\pi ^2}  \Big[ (X -A)  +(Z -B)  \epsilon^2 \Big]
       + O(\epsilon^3)
   \label{regGxi0}
\el
with
\bl
X -A =&
   (m^2-\frac{R}{6}) \Big(\psi ( \frac{3}{2}-\nu)
    +\psi (\frac{3}{2}+\nu) + \ln (\frac{R}{12 m^2}) \Big)
    + \frac{R}{9} \, ,
    \nn \\
Z -B = &   \frac12 m^2 (\frac{R}{6} -m^2)
\Big(  \psi(\frac{3}{2}-\nu)+ \psi(\frac{3}{2}+\nu)
+ \ln \frac{R}{12 m^2} \Big)
+  \frac{R^2 }{48}  -\frac{m^2  R}{18} \,    ,
\label{52}
\el
where a relation
$\psi(\frac52+\nu) +\psi(\frac52-\nu)
= \psi(\frac32+\nu) +\psi(\frac32-\nu)
+\frac14\frac{R}{m^2+\xi R}$   has been used in \eqref{52}.
The expression   \eqref{regGxi0} is valid  only at $m\ne 0$,
as it is derived from the Green's function
\eqref{GreeHyper} which  is undefined at $m=0$ and  $\xi=0$.
If we take massless limit of \eqref{regGxi0},
using $\psi(\frac32+\nu) +\psi(\frac32-\nu)
    \simeq  -\frac{3 H^2}{m^2}$ at small $m$,
we will get the  auto-Green's function
\be \label{addvf}
G(0)_{reg} =  \frac{1}{16\pi ^2} (X -A) \simeq \frac{R^2}{384 \pi^2 m^2}
\ee
which is singular as $m\rightarrow 0$.
The expression \eqref{addvf} and
the conclusion of its invalidity at  $m=0$ and  $\xi=0$
agree with that of Ref.\cite{VilenkinFord1982}.
A singularity at $m=0$ also occurs
in the Green's function of the Proca massive vector field
which does not reduce to the Maxwell field,
and both fields are treated separately
in the Mikowski spacetime \cite{ItzyksonZuber}.
Analogously, we shall give a separate treatment for the case
 $m=\xi=0$ later around \eqref{49} \eqref{59}.
Ref.\cite{VilenkinFord1982} adopted the 4th-order regularized
Green's function of Ref.\cite{BunchDavies1978},
dropping the $R^2$ terms,  and arrived at their (4.10),
which corresponds to
our 2nd-order  auto Green's function $G(0)_{reg}$
given by \eqref{addvf}.
But Ref.\cite{VilenkinFord1982} did not calculate
the regularized stress tensor though,
then continued  to explore
possible quantum states other than the BD vacuum state.
These are beyond the scope of our paper.

The expression \eqref{regGxi0} at small separation
is a simple function of $\epsilon^2$,
and contains neither $\ln\epsilon^2$ nor $\epsilon^2\ln\epsilon^2$ terms.
The stress tensor \eqref{gnt} for $\xi=0$  becomes
\ba\label{TmunuExxi0}
\langle T_{\mu\nu} \rangle_{reg }
&  = &  \lim_{x'\rightarrow x}
\l[  \nabla_ \mu   \nabla_{\nu \, '}    G(x-x')_{reg}
   - \frac12  g_{\mu\nu } \nabla^\sigma   \nabla_{\sigma \, '} G(x-x')_{reg}
   + \frac12  g_{\mu\nu }m^2  G(x-x')_{reg} \r]    .
\ea
Substituting \eqref{regGxi0} into the above
and using the formulae \eqref{D10} and \eqref{limitsigma2} in Appendix,
we obtain  the 2nd-order regularized vacuum stress tensor for $\xi=0$
{\allowdisplaybreaks
\bl
\langle T_{\mu\nu} \rangle_{reg }
& = \frac{1}{16\pi^2}
  \Big[\frac12  g_{\mu\nu }(Z -B )+  \frac12  g_{\mu\nu } m^2 (X -A) \Big]
\nn \\
& = g_{\mu\nu} \Lambda \, ,
\label{Lambda0}
\el
}
where
\be \label{Lambdadef}
\Lambda  \equiv
 \frac{1}{64 \pi^2}
\Big[  m^2 (m^2 - \frac{R}{6})
   \Big(\psi(\frac{3}{2}-\nu) +\psi(\frac{3}{2}+\nu) + \ln \frac{R}{12 m^2} \Big)
  + \frac{m^2 R}{9} + \frac{R^2 }{24} \Big] .
\ee
As we have  checked,
the 2nd-order regularization of
 \eqref{traceTmunu0} also yields the trace of \eqref{Lambda0} consistently.
The stress tensor \eqref{Lambda0}
respects the covariant conservation of energy.
As an important property,
the vacuum stress tensor \eqref{Lambda0} is proportional to the metric,
$ \langle T_{\mu\nu} \rangle_{reg } \propto  g_{\mu\nu} $,
and satisfies the maximal symmetry in de Sitter space.
The finite constant $\Lambda$ of \eqref{Lambdadef}
depends on the mass $m$ of the scalar field
and the expansion rate $H$,
and is naturally identified as, or part of,  the cosmological constant
that drives the de Sitter inflation \cite{Weiberg1983}.
From cosmological point of view,
the cosmological constant
is generally contributed by the vacuum stress tensors
of more than one quantum field.

We compare the results from the point-splitting and from the adiabatic
 for the minimally coupling $\xi=0$.
Fig.\ref{rho2ndregxi0} (b)
plots the 2nd-order point-splitting
$\rho_{reg}$ from \eqref{Lambda0} in the blue line,
and the 2nd-order adiabatic $\rho_{reg}$ from (\ref{intrhoreg})
in the red dots.
The two results are equal over the whole range of $m^2/H^2$,
 positive and finite.
Hence, both adiabatic and point-splitting  regularization
yield the same regularized stress tensor for $\xi=0$.

Next  we calculate the stress tensor by the second scheme
 of the point-splitting method.
Substituting the unregularized Green's function \eqref{grgenr}
into \eqref{unrgrtmunmu} for  $\xi=0$,
using the formulae  in Appendix,
we get
\bl
\langle T_{\mu\nu} \rangle
=& \lim_{x'\rightarrow x} \frac{1}{16 \pi^2}
\Big[  \frac{1}{2\epsilon^4}
(\partial_{\mu}  \partial_{\nu'}+\partial_{\mu'}  \partial_{\nu} )\epsilon^2
-\frac{1}{\epsilon^6} ( \partial_{\nu'}\epsilon^2\cdot \partial_{\mu}\epsilon^2
  +  \partial_{\mu'}\epsilon^2\cdot\partial_{\nu}\epsilon^2 )
\nn \\
& ~~~~ - \frac{W}{2\epsilon^4} (\partial_{\mu'}\epsilon^2\partial_{\nu}\epsilon^2
+\partial_{\mu}\epsilon^2 \partial_{\nu'}\epsilon^2)
+  \frac{W}{2\epsilon^2}
(\partial_{\mu}  \partial_{\nu'}+\partial_{\mu'}  \partial_{\nu} )\epsilon^2
\nn
\\
&~~~~
+\frac12g_{\mu\nu} Y \ln\epsilon^2
+\frac{Y}{2\epsilon^2} (\partial_ \mu\epsilon^2\cdot\partial_{\nu \, '}\epsilon^2
+\partial_ {\mu'}\epsilon^2\cdot\partial_{\nu}\epsilon^2)
\nn
\\
& ~~~~  + \frac12  g_{\mu\nu } \frac{R}{12 \epsilon^2}
+ \frac{1}{2} g_{\mu\nu} \frac{W}{\epsilon^2}
 + (Y+ \frac12 Z)  g_{\mu\nu }
 +\frac12  g_{\mu\nu }m^2
\big(-\frac{1}{\epsilon^2}+ W \ln\epsilon^2  + X \big)
  \Big] .
\label{urt}
\el
The subtraction stress tensor $\langle T_{\mu\nu} \rangle_{sub}$
is obtained by substituting the subtraction Green's function \eqref{subGxi0}
into \eqref{subgrtmunmu},
and has an expression similar to \eqref{urt}
with the replacements $(X,Z) \rightarrow (A,B)$.
Their difference $\langle T_{\mu\nu} \rangle_{reg }
 =\langle T_{\mu\nu} \rangle  -\langle T_{\mu\nu} \rangle_{sub }$
 is the same as the result  \eqref{Lambda0} from the first scheme.

Now consider the case of  $m=0$ and $\xi= 0$,
for which  the formulae  \eqref{GreeHyper} \eqref{regGxi0} do not  apply.
We  directly  start with  the  unregularized Green function
of the minimally-coupling massless scalar field \cite{ZhangYeWang2020,ZhangWangYe2020}
\bl
G(\sigma)
& = -\frac{H^{2}}{8 \pi^{2}}
 \Big[ \frac{1}{\sigma}
+\ln (-\frac{2\tau\tau'}{\tau_0^2} \sigma) \Big]   ,
      \label{49}
\el
where   $\tau_0$ is an arbitrary fixed  constant.
All the terms of \eqref{49} are  UV divergent and should be subtracted off,
and we take $G(\sigma)_{sub} = G(\sigma)$,
so that  the regularized vacuum Green function is zero,
\be   \label{59}
G(\sigma)_{reg}= G(\sigma)- G(\sigma)_{sub}= 0 ,
\ee
and  the regularized stress tensor is also zero,
$\langle T_{\mu\nu}\rangle_{reg}=0$,
the same as  \eqref{m=0conformint}
from adiabatic regularization.
In the second  scheme,
the unregularized stress tensor is
\bl \label{Tmunuxi0un}
\langle T_{\mu\nu}\rangle
&= \frac12\Big[ \nabla_{\mu}\nabla_{\nu'}+\nabla_{\mu'}\nabla_{\nu}
-g_{\mu\nu}\nabla_{\sigma}\nabla^{\sigma'}\Big]   G(x-x')  .
\el
Using the formulae \eqref{D10} through  \eqref{281} in Appendix,
we obtain
\bl
\langle T_{\mu\nu}\rangle
  = & \lim_{x' \rightarrow x}-\frac1{32\pi^2}
  \Big[  -\frac{1}{\epsilon^4}
   (\partial_{\mu}  \partial_{\nu'}+\partial_{\mu'}  \partial_{\nu} )\epsilon^2
  +  \frac{2}{\epsilon^6}(\partial_{\nu'}\epsilon^2\cdot
\partial_{\mu}\epsilon^2
+ \partial_{\nu}\epsilon^2\cdot \partial_{\mu'}\epsilon^2)
\nn \\
& - \frac{R}{6} \frac{1}{\epsilon^4}
 (\partial_{\mu}\epsilon^2 \cdot \partial_{\nu'}\epsilon^2
+\partial_{\mu'}\epsilon^2 \cdot \partial_{\nu}\epsilon^2)
 + \frac{R}{6} \frac{1}{\epsilon^2}   (\partial_{\mu} \partial_{\nu'}
    +\partial_{\mu'} \partial_{\nu} )\epsilon^2
    + g_{\mu\nu}  \frac{R}{12} \frac{1}{ \epsilon^2}
\Big] .
\label{213}
\el
All the terms in \eqref{213} are UV divergent
in the coincidence limit $ x'\rightarrow x$
and should be subtracted off,  so that
$\langle T_{\mu\nu}\rangle_{reg}  =0$,
also agreeing with  \eqref{m=0conformint}.

\begin{figure}[htb]
\centering
\subcaptionbox{}
    {  \includegraphics[width = .45\linewidth]{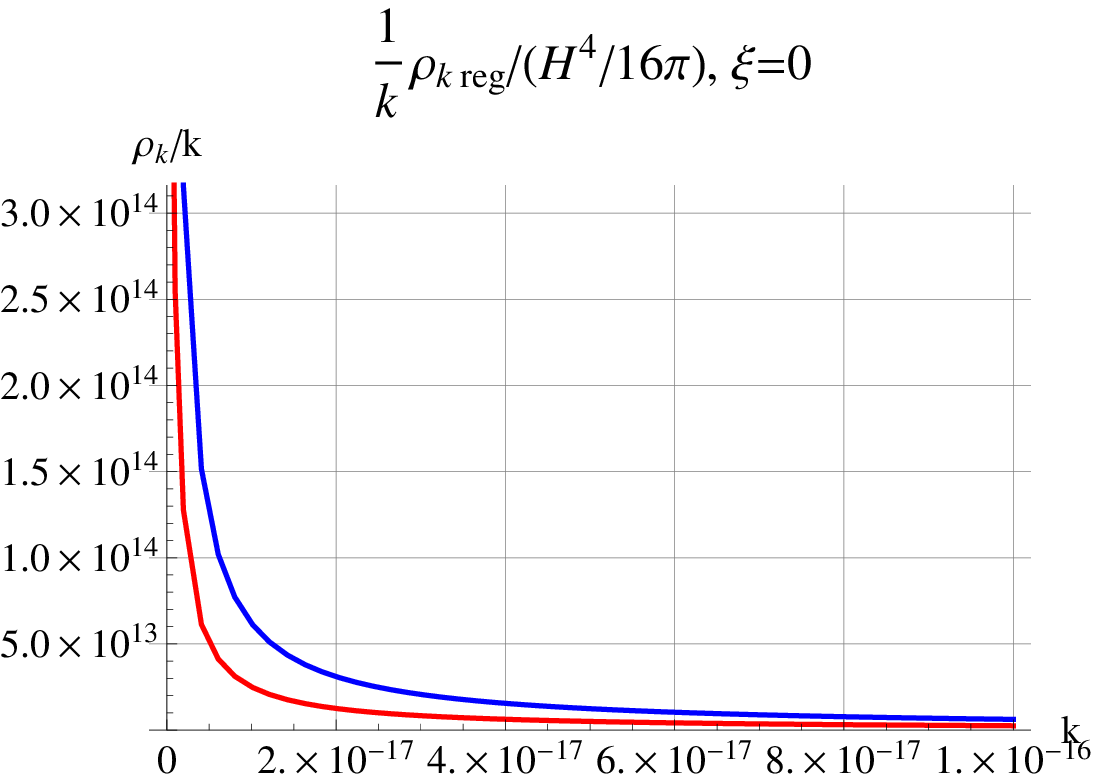}   }
    \subcaptionbox{}
    {  \includegraphics[width = .45\linewidth]{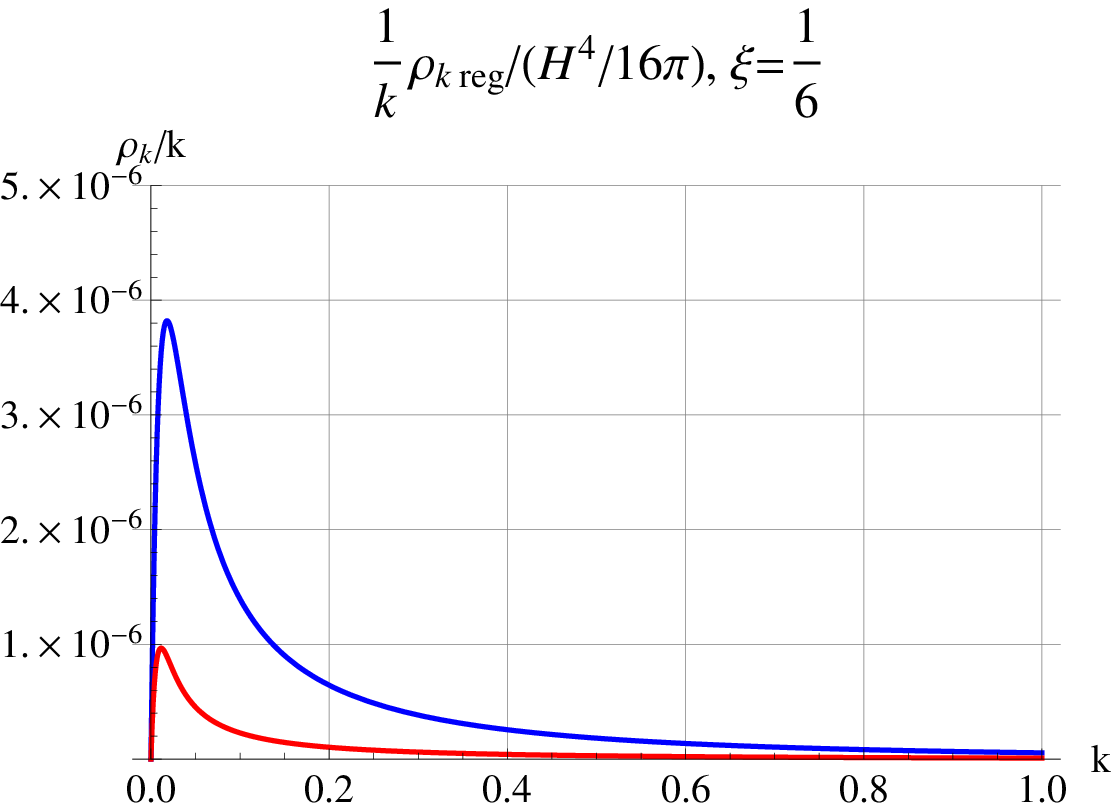}   }
\caption{
(a)  $\frac{1}{k}\rho_{k\, reg}$ for $\xi=0$.
(b)  $\frac{1}{k}\rho_{k\, reg}$ for   $\xi=\frac{1}{6}$.
Red line: $\frac{m^2}{H^2}=4\times 10^{-4}$,
Blue line: $\frac{m^2}{H^2}=10^{-3}$.
}
 \label{xi0smallm}
\end{figure}

As remarked earlier,
the expressions \eqref{regGxi0}  \eqref{addvf} \eqref{Lambda0}  for $\xi=0$
are  valid  only at $m\ne 0$.
If we would take the massless limit of
the stress tensor \eqref{Lambda0},
by the expansion $\psi(\frac32+\nu) +\psi(\frac32-\nu)
    \simeq  -\frac{3 H^2}{m^2} + (\frac{11}{6}-2 \gamma)$ at small $m$,
we would get
\bl \label{limim0}
\lim_{m\rightarrow 0} \langle T_{\mu\nu} \rangle_{reg }
 = g_{\mu\nu }  \frac{1}{64 \pi^2} \Big(  \frac{R^2 }{12} \Big) \ne 0,
\el
in contradiction to the  result  \eqref{m=0conformint} at $m=\xi=0$.
So, our conclusion on the singularity at $m=\xi=0$
is consistent with Ref.\cite{VilenkinFord1982}.
This issue is originated in the $k$-space.
For $\xi=0$, the spectral energy density
$\rho_{k\, reg}$ in \eqref{energykxi0re} has a massless limit
$\lim_{m\rightarrow 0} \rho_{k\, reg} =0$ for any given $k$,
so its $k$-integration is
$\rho_{reg}= \int \lim_{m\rightarrow 0} \rho_{k\, reg} \frac{dk}{k} = 0$.
If we would do the $k$-integration first and then take the massless limit,
we would get the nonvanishing $\rho_{reg}\ne 0$ as \eqref{limim0},
which is actually invalid at $m=0$.
Obviously,  the ordering of the massless limit and the $k$-integration
can not be interchanged
\bl \label{xi=limit}
\lim_{m\rightarrow 0} \int\rho_{k\, reg}\frac{1}{k}dk
\neq
\int \lim_{m\rightarrow 0} \rho_{k\, reg} \frac{1}{k}dk  .
\el
This is because
$\frac{1}{k} \rho_{k\, reg}$ for $\xi=0$ is not dominantly convergent,
ie, there  exists no non-negative integrable function $g_k$
such that  $|\frac{1}{k}\rho_{k\, reg}|\leqslant g_{k}$
for all  $k$ and  $m$.
When $k$ is sufficiently small,
$\frac{1}{k}\rho_{k\, reg}$ is increasingly large,
as shown in Fig.\ref{xi0smallm} (a).

So far,  for $\xi=0$,
in both schemes of the point-splitting, we have been
guided by $G(\sigma)_{sub}$ of \eqref{subGxi0}
from the 2nd-order adiabatic regularization.
Otherwise,   it may not be easy to choose
an appropriate subtraction  stress tensor.
The calculation of stress tensor is straightforward
because the regularized Green's function \eqref{regGxi0}
is a linear function of  $\epsilon^2$.
Nevertheless, for a general $\xi\ne 0$,
the regularized Green's function may contain
a term  $\epsilon^2\ln\epsilon^2$,
and the calculation of stress tensor
 may not be so simple  in  the point-splitting method,
as we shall see in the following sections.

\section{The 2nd-order regularized stress tensor with small $\xi>0$}

For a small $\xi > 0$,
we also use \eqref{energykxi0re} and \eqref{presskreg} for
the 2nd-order adiabatically regularized spectral stress tensor,
and  the $k$-integrations analogous to \eqref{intrhoreg}
give the regularized stress tensor.
(Again,
the 0th-order regularization would not remove all the UV divergences,
and the 4th-order would lead to  a negative energy density.)

We now   calculate the  stress tensor  in the point-splitting method.
The 2nd-order subtraction Green's functions \eqref{sungrxi0}
for general $\xi$ at small separation  is
\bl
 G(\sigma)_{sub}   & =\frac{1}{16 \pi^2}
\Big(-\frac{1}{\epsilon^2}+W\ln\epsilon^2
+C  + D \epsilon^2\ln\epsilon^2
+ E \epsilon^2\Big),\label{RRrunregGgeneralsub}
\el
where the constants are
\bl
C &=(m^2
+R(\xi-\frac{1}{6}))(-1+2\gamma +\ln(- m^2) )+\xi R-\frac{R}{9} ,
\nn
\\
D & =-\frac{1}{2} m^2(m^2 +(\xi-\frac{1}6)R+\xi R) ,
\nn
\\
E &=-\frac{1}{2}m^2(m^2+R(\xi-\frac16)+\xi R)
(-\frac{5}{2}+2\gamma +\ln(- m^2) )
-\frac{5}{12}\frac{m^2R}{6}
-\frac12m^2 \xi R .
\el
The regularized Green's function is  the following difference
\bl
G(\sigma)_{reg} = G(\sigma) -G(\sigma)_{sub}
&=\frac{1}{16 \pi^2}
\Big((X-C)   +(Y-D) \epsilon^2\ln\epsilon^2 +(Z-E) \epsilon^2\Big) ,
 \label{RrregGgeneral2}
\el
where $G(\sigma)$ is the un-regularized Green's functions \eqref{grgenr}.
Notice that
\eqref{RrregGgeneral2}  contains a term $\sim \epsilon^2\ln\epsilon^2$
with a coefficient $(Y-D )=-\frac{1}{2}\xi(\xi-\frac16)R^2$
which is of the 4th-order  and arises from the coupling $\xi R$.
Although $\epsilon^2\ln\epsilon^2$
is continuous and UV convergent at $\epsilon^2=0$,
it will cause a UV divergent term $\sim \ln\epsilon^2$
in the regularized  stress tensor.
Moreover,
when $G_{reg}(\sigma)$ is plugged into  \eqref{gnt}
 to calculate the stress tensor,
some unwanted 4th-order terms $\sim R^2$ due to $G_{sub}(\sigma)$
will come up.
This is because $G_{sub}(\sigma)$ of  \eqref{RRrunregGgeneralsub}
satisfies an inhomogeneous equation as the following
\bl
(\nabla_{\mu}\nabla^{\mu}+m^2+\xi R)G_{sub}(\sigma)
=\frac{1}{16 \pi^2}\Big(\xi(\xi-\frac16)(-1+2\gamma
 +\ln (-m^2\epsilon^2))-\frac{1}{24}+\frac{5\xi}{36}
+\xi^2 \Big)R^2 .
\label{eqsubgr2nd}
\el
So, instead of the first scheme \eqref{gnt},
we shall work with the second scheme,
using \eqref{unrgrtmunmu} and \eqref{subgrtmunmu} in the following.
By calculation,
using the formulae \eqref{D10}--- \eqref{289}  in Appendix,
the un-regularized  stress tensor for general $\xi$
is given by the following
{ \allowdisplaybreaks
\bl
\langle T_{\mu\nu}\rangle
&=\lim_{x' \rightarrow x}\frac{1}{16 \pi^2}\Big[ \frac12(1-2\xi)
\Big(\frac{1}{\epsilon^4}
 (\partial_{\mu}  \partial_{\nu'}+\partial_{\mu'}  \partial_{\nu} )\epsilon^2
-  \frac{2}{\epsilon^6} (\partial_{\nu}\epsilon^2\cdot \partial_{\mu'}\epsilon^2
 +  \partial_{\nu'}\epsilon^2\cdot \partial_{\mu}\epsilon^2)
\nn \\
& + W  \frac{1}{\epsilon^2}
 (\partial_{\mu}  \partial_{\nu'}+\partial_{\mu'}  \partial_{\nu} )\epsilon^2
-W  \frac{1}{\epsilon^4} (\partial_{\mu}\epsilon^2 \partial_{\nu'}\epsilon^2
  + \partial_{\mu'}\epsilon^2 \partial_{\nu}\epsilon^2)
\nn
\\
& +Y\frac{1}{\epsilon^2} \big(\partial_ \mu\epsilon^2 \cdot\partial_{\nu \, '}\epsilon^2
 + \partial_{ \mu'}\epsilon^2\cdot\partial_{\nu} \epsilon^2\big) \Big)
 \nn
\\
& -\xi \Big(\frac{1}{\epsilon^4}
 (\partial_{\mu}  \partial_{\nu}+\partial_{\mu'}  \partial_{\nu'} )\epsilon^2
- \frac{2}{\epsilon^6} ( \partial_{\nu}\epsilon^2\cdot \partial_{\mu}\epsilon^2
   +\partial_{\nu'}\epsilon^2\cdot \partial_{\mu'}\epsilon^2)
\nn \\
& +W  \frac{1}{\epsilon^2}
 (\partial_{\mu}  \partial_{\nu}+\partial_{\mu'}  \partial_{\nu'} )\epsilon^2
  -W\frac{1}{\epsilon^4} (\partial_{\mu}\epsilon^2 \partial_{\nu}\epsilon^2
     + \partial_{\mu'} \epsilon^2\partial_{\nu'}\epsilon^2)
\nn
\\
& +Y\big(\frac{1}{\epsilon^2}\partial_ \mu\epsilon^2\cdot\partial_{\nu}\epsilon^2
+\frac{1}{\epsilon^2}\partial_ {\mu'}\epsilon^2\cdot\partial_{\nu'}\epsilon^2\big)\Big)
\nn \\
& +\xi  \big( \Gamma_{\mu\nu}^{\alpha}\partial_{\alpha}\epsilon^2
+\Gamma_{\mu'\nu'}^{\alpha'}\partial_{\alpha'}\epsilon^2  \big)
\big(\frac{1}{\epsilon^4}+W\frac{1}{\epsilon^2}\big)
\nn
\\
& +g_{\mu\nu}(\frac16(\xi+\frac14)R+\frac{W}{2})\frac{1}{\epsilon^2}
\nn \\
& +\frac12g_{\mu\nu}Y\ln\epsilon^2  +g_{\mu\nu}Y +\frac12g_{\mu\nu}Z
\nn \\
& +\frac12g_{\mu\nu}m^2  (-\frac{1}{\epsilon^2}+W\ln\epsilon^2+X)
\nn
\\
&+g_{\mu\nu}\xi\frac{ RW}{2}-\xi(-g_{\mu\nu}\frac{R}{4})
 (-\frac{1}{\epsilon^2}+W\ln\epsilon^2+X)\Big] ,
\label{Runregstt}
\el
}
which reduces to \eqref{urt} when $\xi=0$.
(Ref.\cite{BunchDavies1978} gave an expression of
$\langle T_{\mu\nu}\rangle$ in their eq.(3.17),
which still contained some direction-dependent splitting vectors.)
The  subtraction stress tensor is obtained
by replacing $(X,Y,Z)$ by $(C,D,E)$ in the above
{  \allowdisplaybreaks
\bl
\langle T_{\mu\nu}\rangle_{sub}
&=\lim_{x' \rightarrow x}\frac{1}{16 \pi^2}\Big[ \frac12(1-2\xi)
\Big( \frac{1}{\epsilon^4}
 (\partial_{\mu}  \partial_{\nu'}+\partial_{\mu'}  \partial_{\nu} )\epsilon^2
- \frac{2}{\epsilon^6} (\partial_{\nu}\epsilon^2\cdot \partial_{\mu'}\epsilon^2
     +  \partial_{\nu'}\epsilon^2\cdot\partial_{\mu}\epsilon^2 )
\nn \\
& + W  \frac{1}{\epsilon^2}
 (\partial_{\mu}  \partial_{\nu'}+\partial_{\mu'}  \partial_{\nu} )\epsilon^2
-W \frac{1}{\epsilon^4} ( \partial_{\mu}\epsilon^2 \partial_{\nu'}\epsilon^2
  + \partial_{\mu'}\epsilon^2 \partial_{\nu}\epsilon^2 )
\nn
\\
& +D \frac{1}{\epsilon^2} \big(\partial_ \mu\epsilon^2\cdot\partial_{\nu \, '} \epsilon^2
  +\partial_{ \mu'}\epsilon^2 \cdot\partial_{\nu}\epsilon^2\big) \Big)
 \nn
\\
& -\xi \Big(\frac{1}{\epsilon^4}
  (\partial_{\mu}  \partial_{\nu}+\partial_{\mu'}  \partial_{\nu'} )\epsilon^2
- \frac{2}{\epsilon^6} ( \partial_{\nu}\epsilon^2\cdot \partial_{\mu}\epsilon^2
+ \partial_{\nu'}\epsilon^2\cdot \partial_{\mu'}\epsilon^2)
\nn \\
& +W  \frac{1}{\epsilon^2}
 (\partial_{\mu}  \partial_{\nu}+\partial_{\mu'}  \partial_{\nu'} )\epsilon^2
-W\frac{1}{\epsilon^4} ( \partial_{\mu}\epsilon^2 \partial_{\nu}\epsilon^2
 + \partial_{\mu'} \epsilon^2 \partial_{\nu'}\epsilon^2)
\nn
\\
& + D \frac{1}{\epsilon^2}\big(\partial_ \mu\epsilon^2\cdot\partial_{\nu}\epsilon^2
+\partial_ {\mu'}\epsilon^2\cdot\partial_{\nu'}\epsilon^2\big)\Big)
\nn \\
& +\xi \big( \Gamma_{\mu\nu}^{\alpha}\partial_{\alpha}\epsilon^2
+\Gamma_{\mu'\nu'}^{\alpha'}\partial_{\alpha'}\epsilon^2 \big)
 \big( \frac{1}{\epsilon^4}+W\frac{1}{\epsilon^2} \big)
\nn
\\
& +g_{\mu\nu}(\frac16(\xi+\frac14)R+\frac{W}{2})\frac{1}{\epsilon^2}
\nn \\
&
+\frac12g_{\mu\nu} D \ln\epsilon^2 +g_{\mu\nu} D +\frac12g_{\mu\nu} E
\nn \\
& +\frac12g_{\mu\nu}m^2 (-\frac{1}{\epsilon^2}+W\ln\epsilon^2+ C)
\nn
\\
&+g_{\mu\nu}\xi\frac{ RW}{2}
  -\xi(-g_{\mu\nu}\frac{R}{4})(-\frac{1}{\epsilon^2}+W\ln\epsilon^2+ C)\Big],
 \label{subwithoutrepla}
\el
}
The expressions \eqref{Runregstt} and \eqref{subwithoutrepla} are lengthy.
But, all $\epsilon^{-4}$ and $\epsilon^{-2}$ divergent terms
will cancel between \eqref{Runregstt} and \eqref{subwithoutrepla},
and will be denoted as $\big(\epsilon^{-4},  \epsilon^{-2} \,  \text{terms} \big)$
for brevity.
The four convergent terms occurring in
\eqref{Runregstt} and \eqref{subwithoutrepla}
will be collectively denoted  as
\bl
P_{\mu\nu} \equiv   (\frac12- 2\xi) \frac{1}{\epsilon^2}( \partial_\mu
  \epsilon^2\cdot\partial_{\nu \, '}\epsilon^2
   + \partial_{\mu'} \epsilon^2\cdot\partial_{\nu}\epsilon^2 )
-\xi \frac{1}{\epsilon^2} ( \partial_ \mu\epsilon^2\cdot\partial_{\nu}\epsilon^2
   + \partial_ {\mu'} \epsilon^2\cdot\partial_{\nu'}\epsilon^2 ) ,
\label{pathdependent}
\el
which  nevertheless depends in
the path of the coincidence limit.
See \eqref{orderlimit}--\eqref{appendix138} in Appendix B.
We write \eqref{Runregstt} and \eqref{subwithoutrepla} briefly
as the following
{\allowdisplaybreaks
\bl
\langle T_{\mu\nu}\rangle
= & \lim_{x' \rightarrow x}\frac{1}{16 \pi^2}\Big[
\Big(\text{$\epsilon^{-4}$,  $\epsilon^{-2}$ terms} \Big)
+    Y P_{\mu\nu}
\nn
\\
& +\frac12g_{\mu\nu} Y \ln\epsilon^2 +g_{\mu\nu} Y +\frac12g_{\mu\nu} Z
  +\frac12g_{\mu\nu}m^2 ( W\ln\epsilon^2+ X)
\nn
\\
& +g_{\mu\nu}\xi\frac{ R}{2} W
  -\xi(-g_{\mu\nu}\frac{R}{4}) W\ln\epsilon^2
  -\xi(-g_{\mu\nu}\frac{R}{4})  X
  \Big] ,
\label{Runregsttst}
\el
}
and
\bl
\langle T_{\mu\nu}\rangle_{sub}
= & \lim_{x'  \rightarrow x}\frac{1}{16 \pi^2}\Big[
\Big(\text{$\epsilon^{-4}$,  $\epsilon^{-2}$ terms} \Big)
 + D P_{\mu\nu}
\nn
\\
& +\frac12g_{\mu\nu} D \ln\epsilon^2 +g_{\mu\nu} D
  +\frac12g_{\mu\nu} E
  +\frac12g_{\mu\nu}m^2 ( W\ln\epsilon^2+ C )
\nn
\\
& +g_{\mu\nu}\xi\frac{ R}{2} W
  -\xi(-g_{\mu\nu}\frac{R}{4}) W\ln\epsilon^2
   -\xi(-g_{\mu\nu}\frac{R}{4}) C
  \Big]  .
 \label{subwithoutreplast}
\el
Recall that, in the 2nd-order adiabatic regularization in $k$-space,
only the 2nd adiabatic order terms $\sim a'\,^2, a'' $,
 are kept in  the subtraction terms $\rho_{k\,A 2}$ and $p_{k\,A 2}$
in \eqref{energykxi0re} and \eqref{presskreg}.
To be consistent, in $x$-space too,
we keep up to the 2nd-order terms
in the subtraction stress tensor \eqref{subwithoutreplast}.
The 4th-order terms $\propto R^2$
come only from the last line of \eqref{subwithoutreplast}:
\bl
R W &= R \Big(m^2+ R (\xi-\frac16)\Big)  =m^2R+(\xi-\frac{1}{6})R^2 ,
  \label{higherorderWR}
\\
R C  &=R \Big((m^2 +R(\xi-\frac{1}{6}))(-1+2\gamma +\ln(- m^2) )
              +\xi R-\frac{R}{9}\Big)
\nn \\
& = R m^2 (-1+2\gamma +\ln(- m^2))
  +R^2 (\xi-\frac{1}{6})(-1+2\gamma +\ln(- m^2))
 +\xi R^2-\frac{R^2}{9} ,
\el
which can be dropped by
 the following replacements in  \eqref{subwithoutreplast},
\bl
R  W   &\rightarrow  R  [ W-(\xi-\frac{1}{6})R  ]  , \label{replaceWR}
\\
R C   &\rightarrow  R [ C -R(\xi-\frac{1}{6})
  (-1+2\gamma +\ln(- m^2) )
   -\xi R+\frac{R}{9}  ]   . \label{replaceOR}
\el
With this replacement, the subtraction stress tensor
\eqref{subwithoutreplast} is modified to the following
{\allowdisplaybreaks
\bl
\langle T_{\mu\nu}\rangle_{sub}
= & \lim_{x'  \rightarrow x}\frac{1}{16 \pi^2}\Big[
\Big(\text{$\epsilon^{-4}$,  $\epsilon^{-2}$ terms} \Big)
 + D P_{\mu\nu}
\nn
\\
& +\frac12g_{\mu\nu} D\ln\epsilon^2
 +g_{\mu\nu} D  +\frac12g_{\mu\nu} E
  +\frac12g_{\mu\nu}m^2 ( W\ln\epsilon^2+ C) \Big]
 \nn \\
&+g_{\mu\nu}\xi\frac{ R}{2}  [W-(\xi-\frac{1}{6})R]
 -\xi(-g_{\mu\nu}\frac{R}{4})   [W-(\xi-\frac{1}{6})R ] \ln\epsilon^2
 \nn \\
&   -\xi(-g_{\mu\nu}\frac{R}{4})[C -R(\xi-\frac{1}{6})(-1+2\gamma +\ln( -m^2) )
   -\xi R+\frac{R}{9}] \Big]  ,
\label{subwithoutrepladropr2}
\el
}
which contains no terms  $\sim R^2$.
We take the difference between \eqref{Runregsttst} and \eqref{subwithoutrepladropr2}
and get
{\allowdisplaybreaks
\bl
\langle T_{\mu\nu}\rangle_{reg}
= &   \frac{(Y- D)}{16 \pi^2}
 \lim_{x' \rightarrow   x} P_{\mu\nu}
+ g_{\mu\nu} \frac{1}{64 \pi^2}  \Big[-3(Y-D )+m^2(X- C)
        -\frac1{4}(\xi-\frac16)R^2\Big] ,
 \label{dropr2}
\el
}
where  the terms $\ln \epsilon^2$ have been  canceled.
However,   the term $\sim \lim P_{\mu\nu}$ in \eqref{dropr2}
depends on the path of the coincidence limit,
and does not possess  the maximum symmetry in de Sitter space.
See \eqref{orderlimit}-- \eqref{appendix138}  in Appendix B.
Dropping the $P_{\mu\nu}$ term from \eqref{dropr2},
we obtain
the 2nd-order regularized vacuum stress tensor with general $\xi$
\bl
\langle T_{\mu\nu}\rangle_{reg}
= &   g_{\mu\nu} \frac{1}{64 \pi^2} \Big[-3(Y-D )+m^2(X- C)
        -\frac1{4}(\xi-\frac16)R^2\Big]
\nn \\
 = &  g_{\mu\nu}  \Lambda
  \label{trace2ndreg2}
\el
where
\bl  \label{Lambda}
  \Lambda & \equiv
 \frac{1}{64 \pi ^2} \bigg[
  m^2 \big(m^2+(\xi-\frac{1}{6})R \big)
\Big(\psi ( \frac{3}{2}-\nu) +\psi ( \frac{3}{2}+\nu)
- \ln (\frac{ 12 m^2}{R}) \Big)
\nn \\
& ~~~~~~~~~~~
-   (\xi-\frac{1}{6})  m^2 R   -\frac{m^2R}{18}
   +\frac{ 3 (\xi-\frac{1}{6})^2 R^2}{2}      \bigg] .
\el
The vacuum stress tensor \eqref{trace2ndreg2}  possesses the maximum symmetry
in de Sitter space.
The constant
 $\Lambda$ of \eqref{Lambda} is identified as the cosmological constant
for $\xi >  0$.
Setting $\xi=0$, \eqref{Lambda} will reduce  to  \eqref{Lambdadef} consistently.

\begin{figure}[htb]
\centering
\subcaptionbox{}
    {  \includegraphics[width = .45\linewidth]{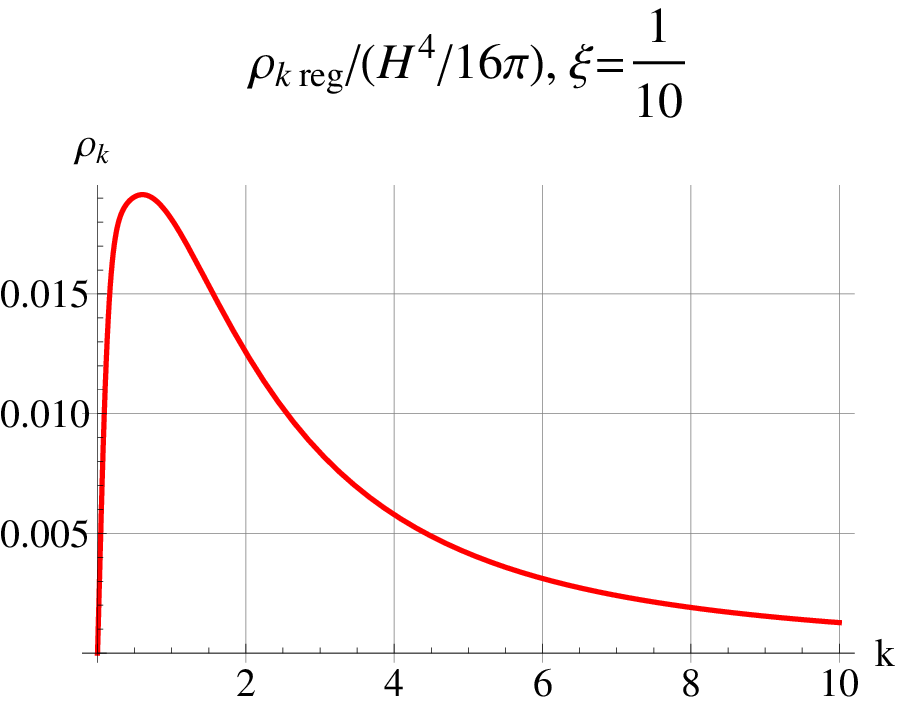}   }
\subcaptionbox{}
    {  \includegraphics[width = .48\linewidth]{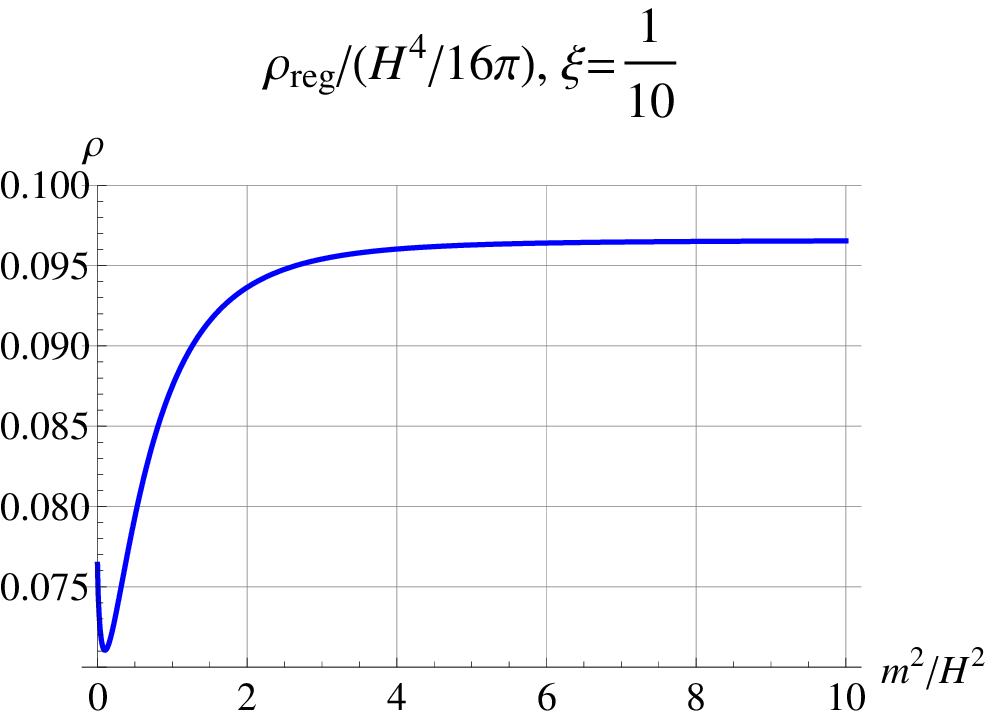}    }
\caption{
(a): The 2nd-order  $\rho_{k\, reg }$
  is positive, IR and UV convergent. The model $\xi=\frac{1}{10}$
     and $\frac{m^2}{H^2}=0.1$.
(b):  For  $\xi=\frac{1}{10}$,   the 2nd-order  $\rho_{reg}$
   is positive and finite for the whole range of  $\frac{m^2}{H^2}$.
}
 \label{rhok2nd110}
\end{figure}

\begin{figure}[htb]
\centering
\subcaptionbox{}
    {  \includegraphics[width = .45\linewidth]{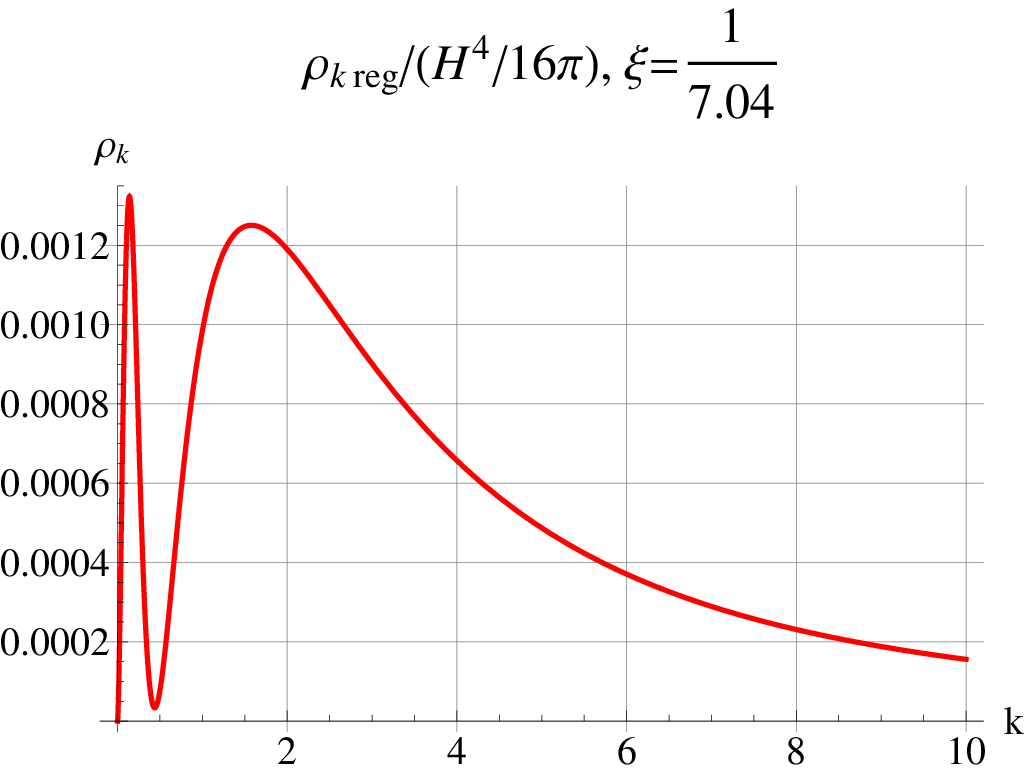}   }
\subcaptionbox{}
    {  \includegraphics[width = .45\linewidth]{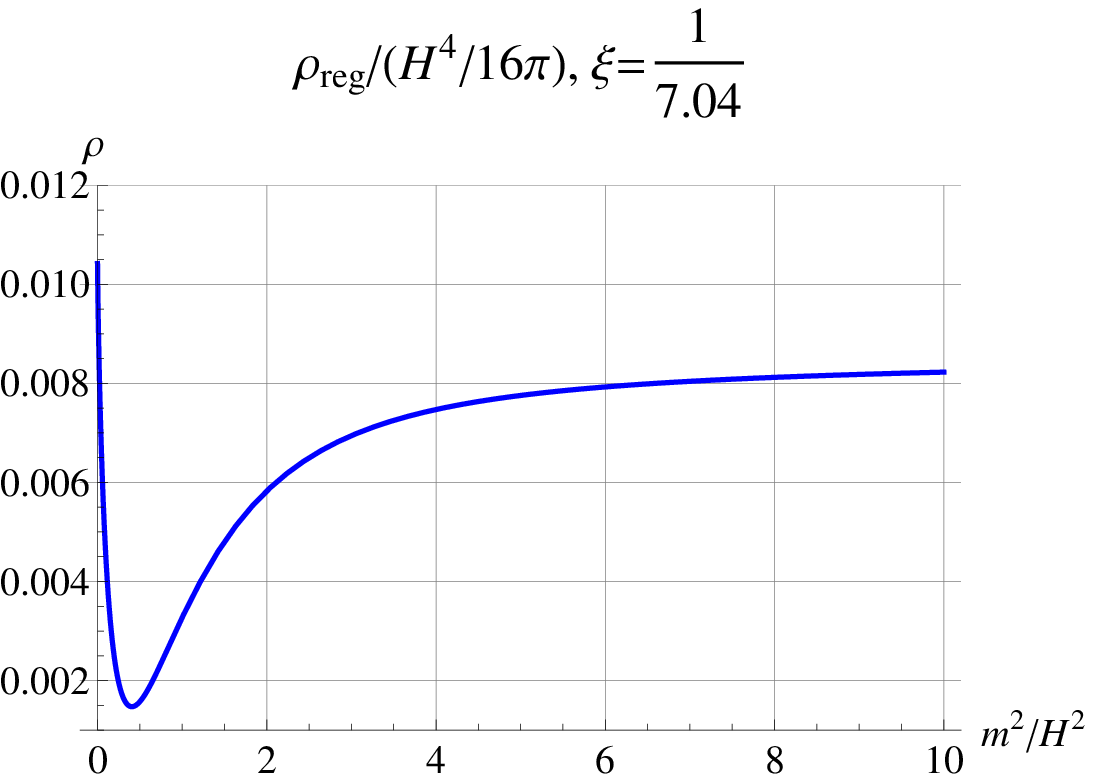}    }
\caption{
(a): The 2nd-order  $\rho_{k\, reg }$
  is positive, IR and UV convergent.
  The model  $\xi=\frac{1}{7.04}$  and $\frac{m^2}{H^2}=0.1$.
(b):  For  $\xi=\frac{1}{7.04}$,   the 2nd-order  $\rho_{reg}$
   is positive and finite for the whole range of  $\frac{m^2}{H^2}$.
}
 \label{rho2nd704}
\end{figure}

It is checked that \eqref{trace2ndreg2}
is equal to  \eqref{intrhoreg} for various  $\xi$ and $m$.
So  the point-splitting and adiabatic regularization of 2nd-order
yield the same result.
Importantly, the 2nd-order regularized
energy density and spectral energy density
are all positive for small couplings  $0 \leq \xi< \frac{1}{7.04}$
at a fixed $\frac{m^2}{H^2}=0.1$.
As  examples,
we plot  $\rho_{k\, reg }$ and $\rho_{reg}$
in Fig.\ref{rhok2nd110} (a) and (b) for $\xi=\frac{1}{10}$,
and  in Fig.\ref{rho2nd704} (a) and (b) for $\xi=\frac{1}{7.04}$.
Nevertheless,
for large couplings $\xi > \frac{1}{7.04}$,
the 2nd-order regularized energy density
and spectral energy density are negative.
(Later we shall see that
the 4th-order regularization also leads to negative
energy density and spectral energy density   for $\xi > \frac{1}{7.04}$.)

The lesson from this section for $\xi\ne 0$ is that,
even though $G(\sigma)_{reg}$ is continuous, as well as  UV and IR convergent,
the point-splitting regularization does not automatically leads to
an appropriate stress tensor.
The coupling $\xi R$  gives rise to $\epsilon^2 \ln \epsilon^2$ in $G(\sigma)_{reg}$,
and causes unwanted higher-order terms in the stress tensor,
as well as some terms depending on the path of the coincidence limit.
These need be treated
in order to give an appropriate stress tensor
which agrees with that from adiabatic regularization.

\section{The  0th-order regularized   stress tensor for $ \xi= \frac16$ }

\begin{figure}[htb]
\centering
\subcaptionbox{}
    {  \includegraphics[width = .45\linewidth]{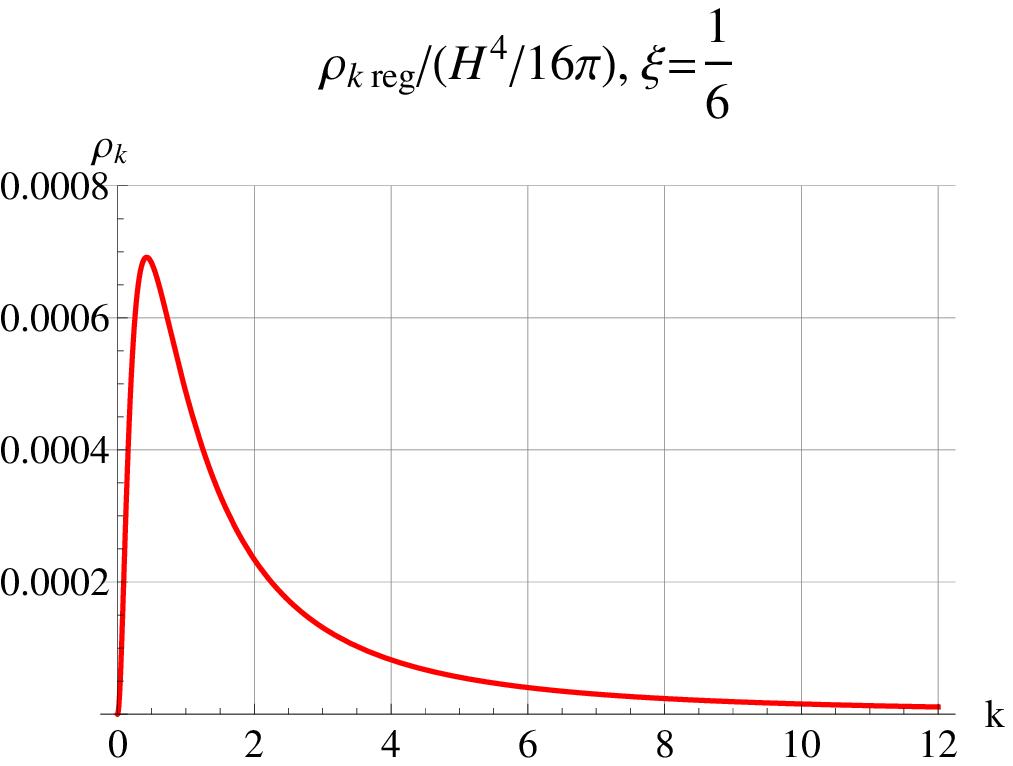}   }
\subcaptionbox{}
    {  \includegraphics[width = .45\linewidth]{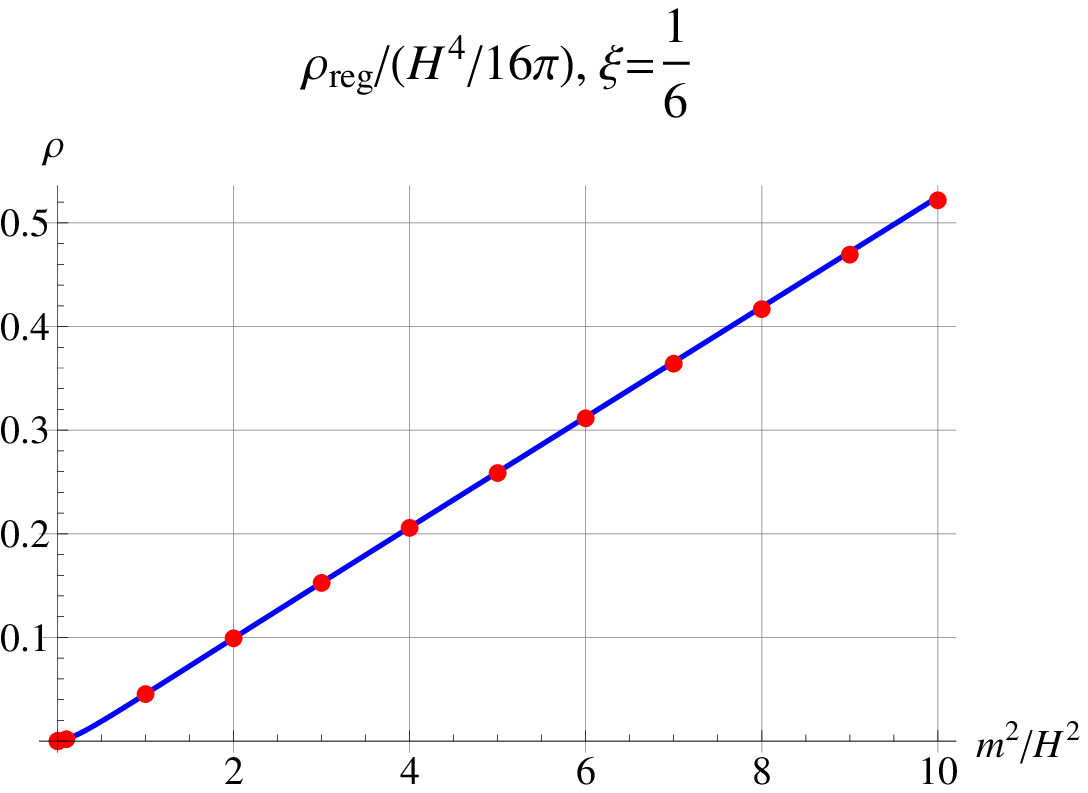}    }
\caption{
(a): The 0th-order  $\rho_{k\, reg }$ in \eqref{rhok16reg}
  is positive, IR and UV convergent.
  The model  $\xi=\frac16 $ and $\frac{m^2}{H^2}=0.1$.
(b):  For $\xi=\frac16$,   the 0th-order $\rho_{reg}$
   is positive and finite for whole range of  $\frac{m^2}{H^2}$.
  Blue line: the point-splitting \eqref{Lambda16};
   Red dots: the adiabatic  \eqref{regrhoxi16}.
}
 \label{rhoxi16}
\end{figure}

We first list the main result
from adiabatic regularization,
and then give the point-splitting regularization.
For a conformally-coupling  $\xi= \frac16$ massive field,
the 0th-order adiabatic regularization is taken on
the spectral stress tensor
\cite{ZhangYeWang2020},
\bl \label{rhok16reg}
\rho_{k\, \, reg}
=& \rho_k - \rho_{k\,A 0}
 \nn \\
= &  \frac{ k^3}{4\pi^2 a^4}
 \Big[ |v_k'|^2 + k^2 |v_k|^2 +m^2 a^2 |v_k|^2 \Big]
      - \frac{ k^3}{4\pi^2 a^4}  \omega ,
\el
\bl
p_{k\, \, reg}
= & p_k - p_{k\,A 0}
\nn \\
=  & \frac{ k^3}{12 \pi^2 a^4}
    \Big[ |v_k'|^2 + k^2|v_k|^2 - m^2 a^2 |v_k|^2 \Big]
       - \frac{k^3}{12 \pi^2 a^4}  \Big(\omega - \frac{m^2 a^2}{\omega} \Big) .
                      \label{pk16reg}
\el
(The 2nd-, and 4th-order regularization
would lead to a negative spectral energy density \cite{ZhangYeWang2020}.)
The 0th-order adiabatically regularized
$\rho_{k\, \, reg}$ and $p_{k\, \, reg}$
are UV and IR convergent,
and $\rho_{k\, \, reg}$ is positive,
as shown in Fig.\ref{rhoxi16} (a).
The 0th-order adiabatically regularized energy density and pressure
 are given by
\be\label{regrhoxi16}
\rho_{reg} =  \int_0^\infty ( \rho_{k} -\rho_{k\, A 0})  \frac{dk}{k} ,
~~~~~
p_{ reg}  =\int_0^\infty ( p_{k} -p_{k\, A 0})  \frac{dk}{k} .
\ee
For examples,
$\rho_{reg} = - p_{ reg} \simeq 0.001786 \frac{H^4}{16 \pi}
            =0.1786  \frac{m^4}{16 \pi} >0$ for $\frac{m^2}{H^2}=0.1$,
and $\rho_{\,reg}=- p_{\, reg} =  0.005221 \frac{H^4}{16 \pi}$
for $\frac{m^2}{H^2}=0.2$.
We plot $\rho_{reg}$ in red dots in Fig.\ref{rhoxi16} (b).
The regularized vacuum stress tensor also satisfies
the maximal symmetry in de Sitter space.
In the massless limit $m=0$ the regularized spectra
and the stress tensor are vanishing
\be\label{rhop16}
\rho_{k\, \, reg} = 0= p_{k\, \, reg} ,
~~~~~ \langle T_{\mu\nu}\rangle_{reg}=0
~~~~~\text{for  $m=0$} ,
\ee
similar to  \eqref{m=0conform} \eqref{m=0conformint} of the case $\xi=0$.

Now we calculate the  stress tensor for $\xi=\frac16$ by the point-splitting method.
The simplest way is
to take  the vacuum expectation of  eq.\eqref{traceTmunu16}
\bl \label{TGrel16}
\langle T^{\mu} \,_{\mu}\rangle_{reg } & = m^2 G(0)_{reg}  ,
\el
and, by the maximal symmetry,
the 0th-order regularized vacuum stress tensor with $\xi=\frac16$
is the following
\bl
\langle T_{\mu\nu }\rangle _{reg }
&=  \frac14 g_{\mu\nu} \langle T^{\alpha } \,_{\alpha  }\rangle_{reg }
=  \frac14 g_{\mu\nu}  m^2 G(0)_{reg} \nn \\
& =   g_{\mu\nu} \Lambda,
 \label{mmx}
\el
where
$G(0)_{reg}$ is the 0th-order
regularized auto-correlation  given by \eqref{G16},
and
\bl\label{Lambda16}
\Lambda \equiv  \frac14   m^2 G(0)_{reg}
   = \frac{m^4 }{64 \pi ^2 } \left[ \psi (\frac{3}{2}-\nu )
      +\psi  (\frac{3}{2}+\nu  )    +  \ln \frac{R}{12 m^2}    \right]
\el
with $\nu = (\frac{1}{4}-\frac{m^2}{H^2})^{1/2}$ for  $\xi=\frac16$.
The merit of this simple derivation is
that no differentiation is performed on the Green's function.
The finite constant of \eqref{Lambda16}
also  can be also identified as the cosmological constant
for the case of conformally-coupling $\xi=\frac16$.

We compare the results from the point-splitting and from the adiabatic
 for the conformally-coupling $\xi=\frac16$.
Fig.\ref{rhoxi16} (b) plots
$\rho_{reg}$ of \eqref{mmx} from the point-splitting in the blue line
 and  $\rho_{reg}$ of \eqref{regrhoxi16} from the adiabatic in the red dots,
the two are equal over the whole range $m^2/H^2$,
positive and finite.
Consider the massless limit of \eqref{mmx}.
By the expansion
$\psi (\frac{3}{2}-\nu ) +\psi (\frac{3}{2}+\nu  )
   \simeq  (1-2 \gamma )+\frac{m^2}{H^2}$ at small $m$,
we have $\Lambda =0$ at $m=0$, so that
\be \label{93}
\langle T_{\mu\nu }\rangle _{reg }=0
~~~~~\text{for  $m=0$} ,
\ee
also agreeing  with \eqref{rhop16}.
Thus, both the point-splitting and adiabatic regularization
to the 0th-order
yield a zero stress tensor
for the conformally-coupling massless scalar field,
and there is no trace anomaly.

Here the ordering of the massless limit
and the $k$-integration of  $\rho_{k\, reg}$  for $\xi=\frac16$
is interchangeable,
\bl
\lim_{m\rightarrow 0}\int\rho_{k\, reg}  \frac{1}{k}dk
  =  \int\lim_{m\rightarrow 0}\rho_{k\, reg}  \frac{1}{k}dk =0  ,
\el
in contrast to the case $\xi=0$ of \eqref{xi=limit}.
This is because  $\frac{1}{k}\rho_{k\, reg}$ satisfies
the requirement of the dominated convergence theorem.
This property  is also reflected by the fact that
the Green's function \eqref{GreeHyper} is valid at $m=0$ and $\xi=\frac16$.
For the illustration,
we plot $\frac{1}{k}\rho_{k\, reg} $
with $\frac{m^2}{H^2}=10^{-3}, 4\times 10^{-3}$ in Fig.\ref{xi0smallm}(b).

Alternatively,  if we apply the formula \eqref{gnt},
\bl\label{tmn16}
\langle T_{\mu\nu}\rangle_{reg}
&=\lim_{x\rightarrow x'}\Big[ \frac13
(\nabla_{\mu}\nabla_{\nu'}+\nabla_{\mu'}\nabla_{\nu})
- \frac16 (\nabla_{\mu}\nabla_{\nu}+\nabla_{\mu'}\nabla_{\nu'} )
- \frac16 g_{\mu\nu}\nabla_{\sigma}\nabla^{\sigma'}\nn
\\
&+\frac16 g_{\mu\nu} ( \nabla_{\sigma}\nabla^{\sigma}
  +\nabla_{\sigma'}\nabla^{\sigma'} )
  -\frac16 G_{\mu\nu} +\frac12m^2g_{\mu\nu}\Big]   G_{reg}(x-x'),
\el
the calculation will be more involved  than that of  eq.\eqref{mmx},
and we shall run into some problems caused by the coupling $\frac16 R$,
similar to the case $\xi > 0$ of Section 5.
The unregularized Green's function  at small separation
is  \eqref{grgenr} with $\xi=\frac16$ and  $W=m^2$,
and the 0th-order  subtraction Green's function \eqref{grsbxi16}
 for  $\xi=\frac16$ at small separation is
\bl
G(\sigma)_{sub}
&=\frac{1}{16\pi^2}
 \Big(-\frac{1}{\epsilon^2}+m^2\ln\epsilon^2
 +K+L\epsilon^2\ln\epsilon^2 +M\epsilon^2 \Big)
   + O(\epsilon^3)   ,\label{299}
\el
with
\bl
 K & = m^2 \Big(-1+2\gamma +\ln(- m^2)  \Big) , \nn
\\
 L & =-\frac{m^4}{2} , \nn
\\
 M & =-\frac{m^4}{4}(-5+4\gamma  +2\ln(- m^2) ) . \nn
\el
So, the difference between \eqref{grgenr} and \eqref{299}
gives the  0th-order regularized Green's function at small distance
{\allowdisplaybreaks
\bl   \label{G16}
G(\sigma )_{reg}
 = & \frac{1}{16 \pi^2} \Big((X -K)+ (Y -L)\epsilon^2\ln\epsilon^2
     +(Z -M)\epsilon^2\Big)
\el
where  $X,Y,Z$ are given in \eqref{X}  \eqref{Y} \eqref{RrZ279},
and
\bl
X -K    & =   m^2 \Big(\ln \frac{R}{12 m^2}
    +\psi(\frac{3}{2}-\nu)+\psi(\frac{3}{2}+\nu)\Big) ,
\nn
\\
Y -L& = -\frac{m^2  R}{12}   ,
\nn
\\
Z -M& =-\frac12 m^4 \Big( \ln \frac{R }{12 m^2}+ \psi(\frac{5}{2}-\nu)
+ \psi(\frac{5}{2}+\nu) \Big), \nn
\\
&~~~   -\frac12 \frac{m^2  R}{12}   \Big( -5+4\gamma   +2\ln(- H^2)
+2 \psi(\frac{5}{2}-\nu)+2 \psi(\frac{5}{2}+\nu) \Big) .
\nn
\el
Note that the term $\epsilon^2\ln \epsilon^2$ appears in \eqref{G16},
like  \eqref{RrregGgeneral2} for general $\xi$.
A calculation shows  that the regularized $G_{reg}(\sigma)$  of \eqref{299}
satisfies the inhomogeneous equation
\be
\big( \nabla^{\mu}\nabla_{\mu}  +\frac16   R  +  m^2  \big) G_{reg}(\sigma )
 =  \frac{1}{16 \pi^2}(Y -L) \Big[ 1+4\gamma+2\ln (-m^2\epsilon^2)\Big] ,
\label{xi16Green}
\ee
which will cause unwanted higher order terms ($\sim R$)
in the stress tensor.
}
Thus, we shall work with the second scheme,
using \eqref{unrgrtmunmu} and \eqref{subgrtmunmu} in the following.
The unregularized stress tensor is \eqref{Runregsttst}
with $\xi=\frac16$ and $W=m^2$,
\bl
\langle T_{\mu\nu}\rangle
= & \lim_{x'  \rightarrow x}\frac{1}{16 \pi^2}\Big[
   \Big(\text{$\epsilon^{-4}$, $\epsilon^{-2}$ terms} \Big)
+Y P_{\mu\nu}
\nn
\\
& +\frac12g_{\mu\nu} Y \ln\epsilon^2 +g_{\mu\nu} Y +\frac12g_{\mu\nu} Z
  +\frac12g_{\mu\nu}m^2 ( m^2 \ln\epsilon^2+ X)
\nn
\\
& +g_{\mu\nu}\xi\frac{ R}{2} m^2
  -\xi(-g_{\mu\nu}\frac{R}{4}) m^2 \ln\epsilon^2
  -\xi(-g_{\mu\nu}\frac{R}{4})  X   \Big]
  ,
\label{16ttst}
\el
and the substraction stress tensor is obtained
by replacing $(X,Y,Z)$ by $( K,L,M)$ in \eqref{16ttst},
\bl
\langle T_{\mu\nu}\rangle_{sub}
= & \lim_{x'  \rightarrow   x  }\frac{1}{16 \pi^2}\Big[
  \Big(\text{$\epsilon^{-4}$, $\epsilon^{-2}$ terms} \Big)
  + L P_{\mu\nu}
\nn
\\
& +\frac12g_{\mu\nu} L \ln\epsilon^2 +g_{\mu\nu} L  +\frac12g_{\mu\nu} M
  +\frac12g_{\mu\nu}m^2 ( m^2 \ln\epsilon^2+ K )
\nn
\\
& +g_{\mu\nu}\xi\frac{ R}{2} m^2
  -\xi(-g_{\mu\nu}\frac{R}{4}) m^2 \ln\epsilon^2
  -\xi(-g_{\mu\nu}\frac{R}{4})  K   \Big] ,
\label{16subttst}
\el
with $\xi=\frac16$.
The last three terms in the above are of the 2nd-order $\sim R$,
and should be dropped, leading to
\bl
\langle T_{\mu\nu}\rangle_{sub}
= & \lim_{x'  \rightarrow x}\frac{1}{16 \pi^2}\Big[
   \Big(\text{$\epsilon^{-4}$, $\epsilon^{-2}$ terms} \Big)
+ L P_{\mu\nu}
\nn
\\
& +\frac12g_{\mu\nu} L \ln\epsilon^2 +g_{\mu\nu} L  +\frac12g_{\mu\nu} M
  +\frac12g_{\mu\nu}m^2 ( m^2 \ln\epsilon^2+ K )
    \Big] ,
\label{16subttst}
\el
where the coefficients $(K,L,M)$ are the 0th-order.
Now the difference between  \eqref{16ttst} and \eqref{16subttst} yields
\bl
\langle T_{\mu\nu}\rangle_{reg}
= &  \frac{ ( Y-L)}{16 \pi^2}
  \lim_{x'   \rightarrow x} P_{\mu\nu}
  +  g_{\mu\nu} \frac{1}{32 \pi^2} \Big[   (Z-M)
     +  m^2  (X-K)  +\frac{R}{12}  X \Big] ,
\label{reg16ttst2}
\el
The  term $\sim \lim P_{\mu\nu}$  is of the 2nd-order,
depends on the path  of the coincidence limit,
and does not possess   the maximum symmetry.
Dropping it, we   obtain
\bl
\langle T_{\mu\nu}\rangle_{reg}
= &   g_{\mu\nu} \frac{1}{32 \pi^2}
\Big[  (Z-M)   +  m^2  (X-K)  +\frac{R}{12}  X \Big] ,
\label{reg16ttst3}
\el
which is equal to the result \eqref{mmx} from the simple derivation.

For the case of $m=0$ and $\xi=\frac16$,
the  unregularized Green function \eqref{GreeHyper} is valid and
reduces to the following simple form \cite{ZhangYeWang2020,ZhangWangYe2020}
\bl
G(\sigma) =-\frac{H^{2}}{8 \pi^{2}}  \, \frac{1}{\sigma} ,
   \label{cfm49}
\el
consisting of one divergent term only.
After subtraction of this term,
the regularized Green's function is  $G(x,x')_{reg} =0$.
This result agrees  with \eqref{G16} at $m=0$.
So, $\langle T_{\mu\nu}\rangle_{reg} = 0$,
also agreeing with \eqref{93} from the adiabatic regularization.
In the second  scheme,  the unregularized stress tensor is
\bl\label{cfmunrgrtmunmu}
\langle T_{\mu\nu} \rangle
 = & \lim_{x'  \rightarrow x  } \Big[ \frac13
     ( \nabla_{\mu}\nabla_{\nu'}+\nabla_{\mu'}\nabla_{\nu} )
       -\frac16  ( \nabla_{\mu}\nabla_{\nu}+\nabla_{\mu'}\nabla_{\nu'} )
       -\frac16  g_{\mu\nu}\nabla_{\sigma}\nabla^{\sigma'}
       \nn   \\
&   + \frac16 g_{\mu\nu} ( \nabla_{\sigma}\nabla^{\sigma}
  +\nabla_{\sigma'}\nabla^{\sigma'} )
   - \frac16   G_{\mu\nu}  \Big]   G(x-x')
\nn \\
=&  -\frac{1}{48\pi^2} \lim_{x' \rightarrow x} \Big[
- \frac{1}{\epsilon^4}
 (\partial_{\mu}  \partial_{\nu'}+\partial_{\mu'}  \partial_{\nu} )\epsilon^2
 +  \frac{1}{2 \epsilon^4}
 (\partial_{\mu}  \partial_{\nu}+\partial_{\mu'}  \partial_{\nu'} )\epsilon^2
 \nn \\
&  +\frac{2}{\epsilon^6} (\partial_{\nu'}\epsilon^2 \cdot \partial_{\mu}\epsilon^2
     +   \partial_{\nu}\epsilon^2 \cdot \partial_{\mu'}\epsilon^2)
 - \frac{1}{\epsilon^6} (\partial_{\nu}\epsilon^2\cdot \partial_{\mu}\epsilon^2
          +   \partial_{\nu'}\epsilon^2\cdot \partial_{\mu'}\epsilon^2 )
\nn \\
& -\frac{1}{12} g_{\mu\nu}R\frac{1}{\epsilon^2}-
\frac{1}{2\epsilon^4}\Gamma_{\mu'\nu'}^{\alpha'}\partial_{\alpha'}\epsilon^{2}
-
\frac{1}{2\epsilon^4}\Gamma_{\mu\nu}^{\alpha}\partial_{\alpha}\epsilon^{2}\Big] .
\el
All the terms in \eqref{cfmunrgrtmunmu} are UV divergent
and should be subtracted off,
we also arrive at   $\langle T_{\mu\nu}\rangle_{reg}  =0$,
the same as  \eqref{93}.

\section{The impropriate  4th-order regularization}

\begin{figure}[htb]
\centering
\subcaptionbox{}
    {  \includegraphics[width = .45\linewidth]{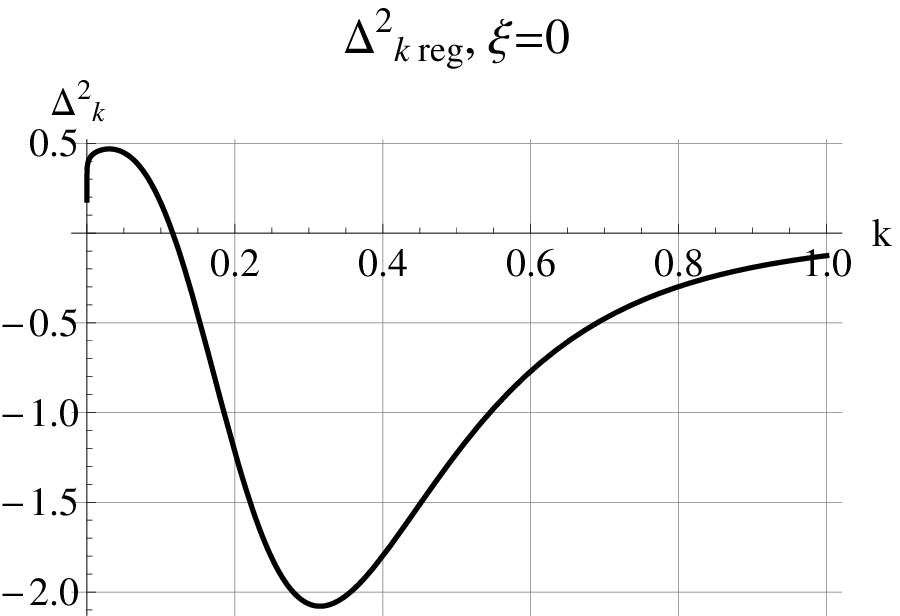}   }
\subcaptionbox{}
    {  \includegraphics[width = .45\linewidth]{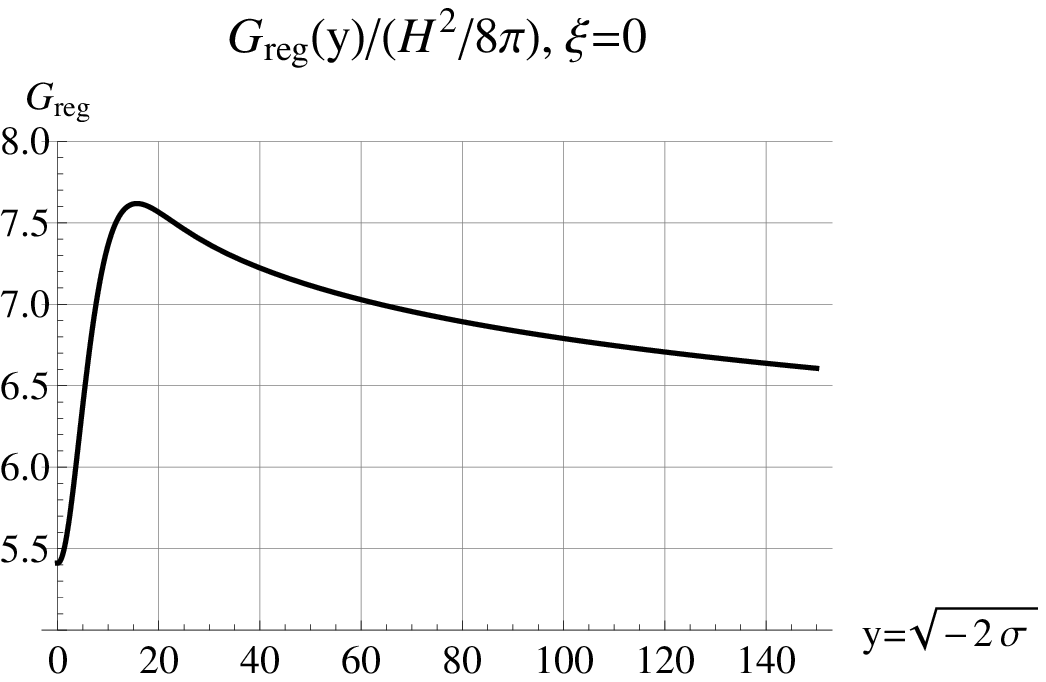}    }
\caption{
(a):  The 4th-order  $\Delta^2_{k\, reg}$ is  negative.
(b):  the 4th-order  $G_{reg}$.
        The model   $\xi=0$  and $\frac{m^2}{H^2}=0.1$.
}
 \label{4thregularizeG}
\end{figure}

\begin{figure}[htb]
\centering
\subcaptionbox{}
    {  \includegraphics[width = .45\linewidth]{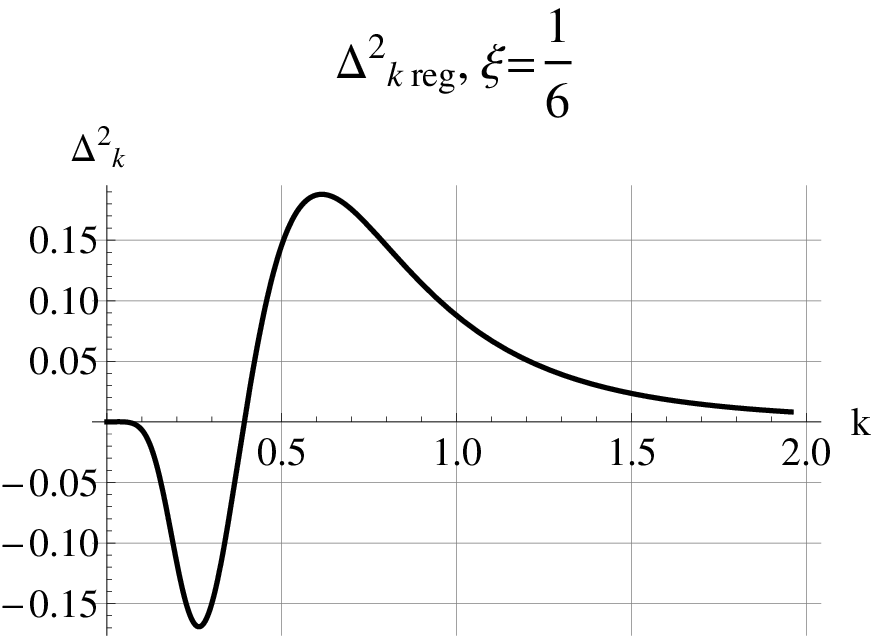}   }
\subcaptionbox{}
    {  \includegraphics[width = .48\linewidth]{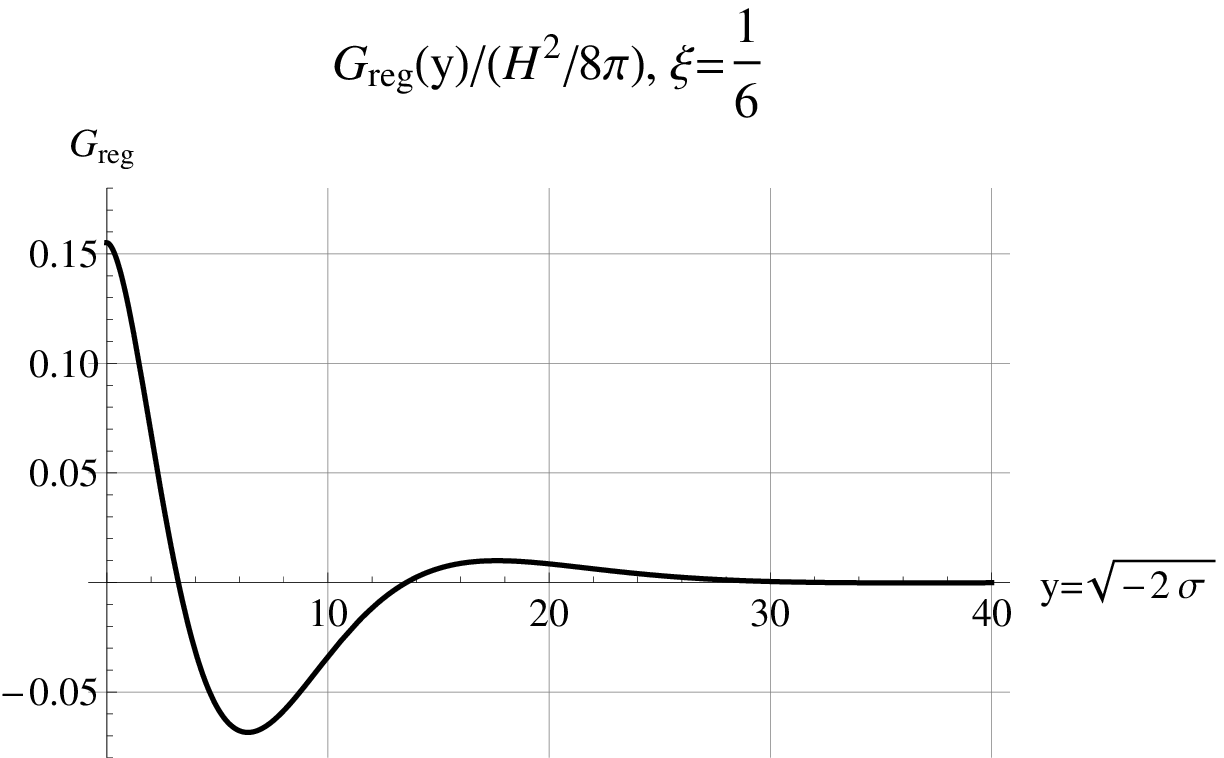}    }
\caption{
(a):  The 4th-order  $\Delta^2_{k\, reg}$ takes  negative values.
(b):  the 4th-order $G_{reg}$.
        The model $\xi=\frac16$  and $\frac{m^2}{H^2}=0.1$.
}
 \label{Greenfunctionreg4th}
\end{figure}

We now examine the conventional 4th-order regularization
for the scalar field with a general $\xi$,
and reveal its  unphysical consequences.
The 4th-order adiabatically regularized power spectrum
 with a general $\xi$ is
\be
\Delta^2_{k\, reg}   =  \frac{ k^{3}}{2  \pi^2 a^2 }
      \Big( |v_k |^2 -\frac{1}{2 W^{(4)}} \Big) ,
\label{4thPS}
\ee
where  the 4th-order effective inverse frequency is
 (see  (a38) in Ref.\cite{ZhangYeWang2020})
\ba\label{W4gen}
(W_k^{(4)})^{-1} & =  &
\frac{1}{\omega} -3 (\xi -\frac{1}{6} )  \frac{1}{\omega ^3} \frac{a''}{a}
  +\frac{m^2  (a a''+ a'\, ^2 )}{4 \omega ^5}
  -\frac{5 m^4 a^2  a'\, ^2}{8 \omega^7}
   \nn \\
&&  -\frac{m^2  (3 a''\, ^2 + a''''  a + 4 a''' a' )}{16 \omega ^7}
    +\frac{7 m^4 \left(3 a^2 a''\, ^2 +3  a'\, ^4+18 a  a'\, ^2 a''
        +4 a^2 a'''  a'\right)}{32 \omega ^9} \nn \\
&& -\frac{231  m^6 a^2 \left( a'\, ^4+a  a'\, ^2 a''\right)}{32 \omega ^{11}}
   +\frac{1155 m^8 a^4   a'\, ^4}{128 \omega ^{13}} \nn \\
&& +  (\xi -\frac{1}{6} )
  \Big[ \frac{3}{4 \omega ^5} ( -\frac{a''\, ^2}{a^2}  + \frac{a'''' }{a}
  +2 \frac{ a'\, ^2 a''}{a^3}  -2 \frac{a'''  a'}{a^2})
     \nn \\
&& -\frac{15 m^2   \left( a''\, ^2 +a'''  a'\right)}{4 \omega ^7}
    +\frac{105  m^4 a  a'\, ^2 a''}{8 \omega ^9} \Big]
    + (\xi -\frac{1}{6} )^2 \, \frac{27}{2\omega ^5} \frac{a''\,^2}{a^2}  .
\ea
The 4th-order power spectrum $\Delta^2_{k\, reg}$   is negative,
as shown in Fig.\ref{4thregularizeG} (a)  for $\xi=0$,
and in Fig.\ref{Greenfunctionreg4th} (a)  for $\xi=\frac16$.
The  negative power spectrum is unphysical.
Obviously,
the 4th-order regularization has subtracted off too much for the scalar field,
and is discordant with the minimum subtraction rule
\cite{ParkerFulling1974}.
Moreover, the 4th-order regularization will cause
other difficulties,  as we shall examine  in the following.

The 4th-order subtraction Green's function is given by
\bl \label{sungrxi04th}
G(y)_{sub} & =  \frac{H^2}{8\pi} \frac{1}{y} \int_0^\infty d k
  \,  k \sin( k y) \Big(    \frac{2}{\pi}\frac{1}{ W^{(4)}} \Big)
\nn \\
& =  \frac{H^2}{4\pi^2} \Bigg[
       \frac{m}{H}  \, \frac{1}{y} K_1 \Big( \frac{m}{H}  y \Big)
\nn
\\
&   -6(\xi -\frac16) K_0 \Big( \frac{m}{H}  y \Big)  +  \frac{1}{4}
       \frac{m}{H}  \, y  K_1 \Big( \frac{m}{H}  y \Big)
     -  \frac{1}{ 24 } \frac{m^2}{H^2} y^2  K_2 \Big( \frac{m}{H}  y \Big)
     \nn \\
&  + \frac{ 9  (\xi -\frac{1}{6} ) +54 (\xi -\frac{1}{6})^2 }{3}
     (\frac{m}{H})^{-1}  \, y  K_1 \Big( \frac{m}{H} y \Big)
       -\frac{1 +10  (\xi -\frac{1}{6} )}{4}    y^2  K_2 \Big( \frac{m}{H} y \Big)
       \nn \\
&  +  \frac{525+840  (\xi -\frac{1}{6})}{32\cdot 105} (\frac{m}{H})
         y^3  K_3 \Big(\frac{m}{H} y \Big)   -\frac{693 }{32 \cdot 945 }
         (\frac{m}{H})^{2}  y^4  K_4 \Big(\frac{m}{H} y \Big)
         \nn \\
&  +  \frac{1155 }{128 \cdot 10395 }
       (\frac{m}{H})^{3}  y^5  K_5 \Big(\frac{m}{H} y \Big)   \Bigg] .
\el
This  4th-order subtraction Green's function
has not been given before in literature.
The first line in \eqref{sungrxi04th}
is the 0th-order subtraction term,
 the first two lines belong to the 2nd-order  subtraction term,
and the remaining terms come from the 4th-order.
Replacing  $y \rightarrow  \sqrt{-2\sigma}$ in \eqref{sungrxi04th}
 gives  $G(\sigma)_{sub}$ for general spacetime separation $\sigma$.
The 4th-order regularized Green's function is given by
\bl\label{Greenreg4}
G(\sigma )_{reg}
& =   G(\sigma )  -G(\sigma)_{sub},
\el
where   $ G(\sigma)$ is given by eq.(\ref{GreeHyper}).
We plot  $G(\sigma)_{reg}$
in Fig.\ref{4thregularizeG} (b) and Fig.\ref{Greenfunctionreg4th} (b).

\begin{figure}[htb]
\centering
\subcaptionbox{}
    {  \includegraphics[width = .45\linewidth]{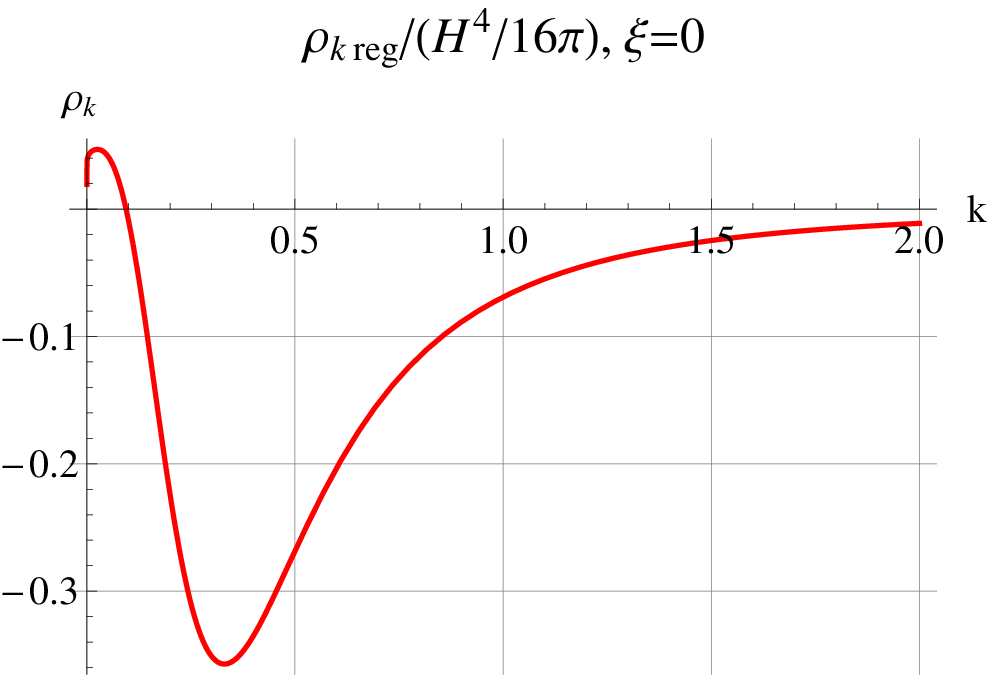}   }
\subcaptionbox{}
    {  \includegraphics[width = .45\linewidth]{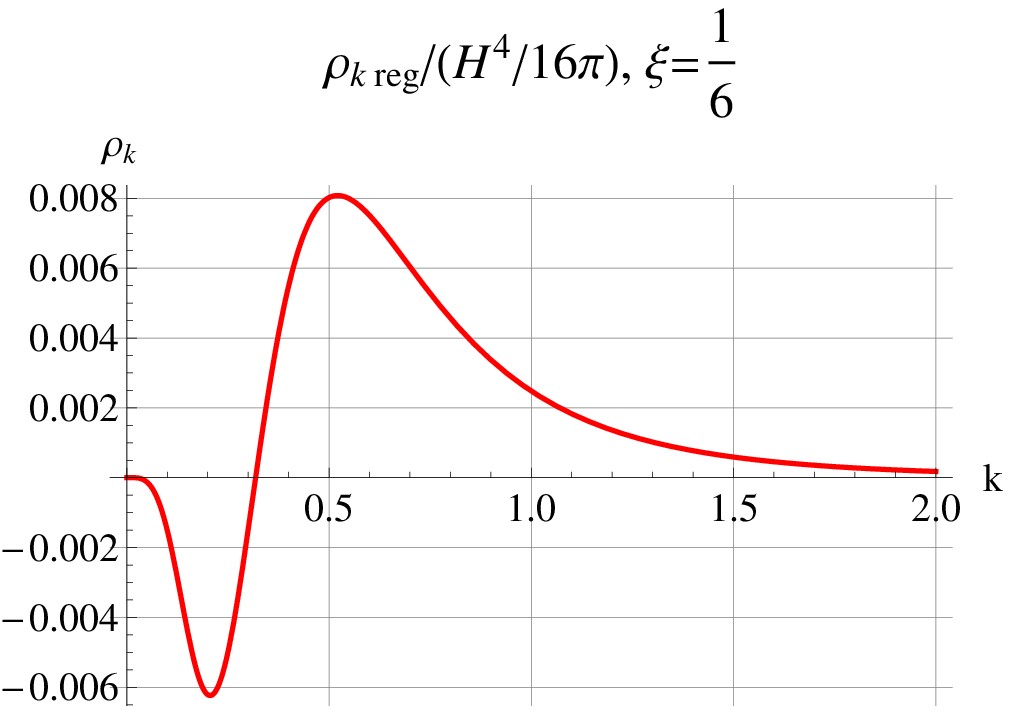}    }
\caption{The 4th-order $\rho_{k\, reg}$ takes negative values.
(a): $\xi=0$.
(b):  $\xi=\frac16$. The model $\frac{m^2}{H^2}=0.1$.
}
 \label{4thregulenerg16}
\end{figure}

As has been found in Ref.\cite{ZhangYeWang2020},
the 4th-order adiabatically  regularized  spectral energy density
$\rho_{k\, reg} = \rho_k - \rho_{k\,A 4}$  takes  negative values too,
as illustrated in Fig.\ref{4thregulenerg16} for $\xi=0$ and $\xi=\frac16$.

We now  calculate the  stress tensor  with general $\xi$
by the 4th-order regularization in the point-splitting method.
At small separation, the 4th-order subtraction Green's function   is
\bl
G(\sigma )_{sub}&=\frac{1}{16\pi^2}
\Big[ -\frac{1}{\epsilon^2 }
+W\ln\epsilon^2
+V + Y\epsilon^2\ln\epsilon^2+T\epsilon^2 \Big]
+ O(\epsilon^3)  ,
\label{RRGreen290}
\el
where
\bl
V= & \Big( m^2+ R (\xi-\frac16) \Big)
    \Big( 2 \gamma-1 +\ln(- m^2) \Big)
+\frac{R}{18} + (\xi-\frac16) R
\nn \\
& -\frac{R^2}{2160m^2} +\frac{(\xi-\frac16)^2R^2}{2m^2} ,
\nn \\
T= & - \frac12(m^2+\xi R)(m^2+(\xi-\frac16)R)
\Big(-\frac52+2 \gamma +\ln(- m^2)
+\frac14\frac{R}{ m^2+ \xi R} \Big)
\nn
\\
&  -\Big(\frac{m^2(\xi-\frac16) R}{2}
  +\frac{m^2R}{36}+\frac{3(\xi-\frac16)^2 R^2}{4}
  +\frac{19R^2}{4320}+\frac{(\xi-\frac16)R^2}{9}\Big).
  \nn
\el
The difference between   \eqref{grgenr} and \eqref{RRGreen290}
is the 4th-order regularized Green's function at small separation
\be
G(\sigma )_{reg}   =   G(0)_{reg}   + \epsilon ^2 G_{\epsilon}  ,
\label{RGreenxi}
\ee
where
\bl
G(0)_{reg}   \equiv & \frac{1}{16\pi^2}(X-V)
\nn \\
= & \frac{1}{16\pi ^2}  \Big[
 \big( m^2+ (\xi-\frac{1}{6} )R \big)
 \Big( \psi(\frac{3}{2}-\nu )+\psi(\frac{3}{2}+\nu )
    - \ln \frac{12 m^2}{R}  \Big)
\nn \\
&  -  (\xi-\frac{1}{6} ) R
  -\frac{1}{18} R
   -\frac{R^2\left(\xi-\frac{1}{6}\right)^2}{2 m^2}
  +\frac{R^2}{2160 m^2 }  \Big] ,  \label{G04t}
  \\
G_{\epsilon}   \equiv & \frac{1}{16\pi^2} (Z-T) .
\label{108}
\el
Our 4th-order  \eqref{RGreenxi} with \eqref{G04t} \eqref{108}
at small separation is equal to (3.14) of
Ref.\cite{BunchDavies1978},
which did not give the subtraction Green's function \eqref{sungrxi04th}
valid for the whole range of $\sigma$.
Since   \eqref{RGreenxi} contains no $\epsilon^2 \ln\epsilon^2$ term,
we do calculation  in the first scheme.
Plugging  \eqref{RGreenxi} into \eqref{gnt}
leads to the following stress tensor,
\bl
\langle T_{\mu\nu}\rangle_{reg}
= & g_{\mu\nu} \Big[  \frac12 G_{\epsilon}
  +\frac12 m^2 G(0)_{reg}   + \frac14\xi R G(0)_{reg} \Big],
  \label{tmunuregg}
\el
which is independent of the path of coincidence limit,
but still contains some unwanted 6th-order terms $\sim R^3$.
This is because $G_{sub}(\sigma)$ of \eqref{RRGreen290}
satisfies the   inhomogeneous equation
\bl
\lim_{x' \rightarrow x  } (\nabla_{\sigma}\nabla^{\sigma}+m^2+\xi R)G_{sub}(\sigma)
 =  \frac{1}{16 \pi^2}\Big(-\frac{\xi}{2160m^2}
      +\frac{\xi(\xi-\frac16)^2}{2m^2}  \Big)R^3  ,
  \label{sub4thgeneralxi0}
\el
due to the coupling $\xi R$.
Requiring the 4th-order Green's function to
satisfy the homogeneous equation to the 4th-order,
\be
\lim_{x' \rightarrow    x}\Big[  \nabla^{\mu}\nabla_{\mu} +\xi R
    + m^2 \Big] G_{reg}(\sigma)  =  0 ,
 \label{4theq}
\ee
ie,
\be
\xi R   G_{reg}(0)
 =  -  \lim_{x' \rightarrow x}\Big[  \nabla^{\mu}\nabla_{\mu}
        + m^2 \Big] G_{reg}(\sigma )
         =  - 2G_{\epsilon} - m^2 G_{reg}(0).
\ee
By this relation,
we can replace $\xi R G_{reg}(0)$ in   \eqref{tmunuregg}
by $(- 2G_{\epsilon} - m^2 G_{reg}(0))$,  and arrive at
the 4th-order regularized stress tensor
\bl
\langle T_{\mu\nu}\rangle_{reg}
= &   g_{\mu\nu}  \frac14   m^2 G_{reg}(0)
 \nn \\
= &   g_{\mu\nu}  \frac{1}{64\pi ^2}
\Big[  m^2(m^2+(\xi-\frac{1}{6})R)
   \Big( \psi(\frac{3}{2}-\nu )+\psi(\frac{3}{2}+\nu) + \ln \frac{R}{12 m^2} \Big)
   \nn \\
&  ~~~~~~~  -  m^2(\xi-\frac{1}{6})R    -\frac{m^2R}{18}
   -\frac{(\xi-\frac{1}{6})^2  R^2 }{2}   +\frac{R^2}{2160 }  \Big] ,
 \label{trted4thxitmunureg}
\el
containing  no $R^3$ terms.

\eqref{trted4thxitmunureg} can be also derived by the second scheme.
The subtraction stress tensor is obtained by replacing  $(X,Z)$ by $(V,T)$,
\bl
\langle T_{\mu\nu}\rangle_{sub}
= & \lim_{x'  \rightarrow x}\frac{1}{16 \pi^2}\Big[
\Big(\text{$\epsilon^{-4}$,  $\epsilon^{-2}$ terms} \Big)
  + Y P_{\mu\nu}
\nn
\\
& +\frac12g_{\mu\nu} Y \ln\epsilon^2 +g_{\mu\nu} Y +\frac12g_{\mu\nu} T
  +\frac12g_{\mu\nu}m^2 ( W\ln\epsilon^2+ V)
\nn
\\
& +g_{\mu\nu}\xi\frac{ R}{2} W
  -\xi(-g_{\mu\nu}\frac{R}{4}) W\ln\epsilon^2
  -\xi(-g_{\mu\nu}\frac{R}{4})  V
  \Big] .
\label{substtst}
\el
The last term $R V$ of \eqref{substtst} contains  $R^{3}$
which can be dropped by the replacement
\[
R V \rightarrow R \Big(V +\frac{R^2}{2160m^2} -\frac{(\xi-\frac16)^2R^2}{2m^2} \Big),
\]
yielding
\bl
\langle T_{\mu\nu}\rangle_{sub}
= & \lim_{x' \rightarrow   x }\frac{1}{16 \pi^2}\Big[
\Big(\text{$\epsilon^{-4}$,  $\epsilon^{-2}$ terms} \Big)
+ Y P_{\mu\nu}
\nn
\\
& +\frac12g_{\mu\nu} Y \ln\epsilon^2 +g_{\mu\nu} Y +\frac12g_{\mu\nu} T
  +\frac12g_{\mu\nu}m^2 (W\ln\epsilon^2 + V)
\nn
\\
& +g_{\mu\nu}\xi\frac{ R}{2} W
  -\xi(-g_{\mu\nu}\frac{R}{4}) W\ln\epsilon^2
  -\xi(-g_{\mu\nu}\frac{R}{4})
  \Big(V +\frac{R^2}{2160m^2} - \frac{(\xi-\frac16)^2R^2}{2m^2}  \Big)
  \Big] .
\label{modsubsttst}
\el
The difference between \eqref{Runregsttst} and \eqref{modsubsttst} yields
the 4th-order regularized vacuum stress tensor
\bl
\langle T_{\mu\nu}\rangle_{reg}
= & g_{\mu\nu} \frac{1}{64  \pi^2}  \Big[
    2 (Z - T)  +2  m^2 (X-V)
 +  \xi   R (X- V) -  \frac{\xi  R^3}{2160m^2}
      +  \frac{\xi  (\xi-\frac16)^2 R^3}{2m^2}     \Big] ,
\label{regregsttst}
\el
which is equal to  \eqref{trted4thxitmunureg}.

Now we examine several difficulties associated with
the outcome of  4th-order regularization.
Firstly   the last two terms of  $G(0)_{reg}$  in \eqref{G04t}
are proportional to $m^{-2}$ and singular at $m=0$,
so that  the 4th-order regularized Green's function \eqref{RGreenxi}
is ill-defined in the  massless limit.
Associated with this
is the so-called trace anomaly for $\xi=\frac16$
in the massless limit \cite{BunchDavies1978},
\bl  \label{trgenxtmnnew}
\lim_{m=0}
\langle T^{\mu}\, _{\mu}\rangle_{reg}
   = m^2 G(0)_{reg}  =  \frac{R^2}{34560\pi^2} ,
\el
which comes exactly from the last, singular term $\frac{R^2}{34560m^2\pi^2}$
in  \eqref{G04t}.
Obviously,
the 4th-order result \eqref{trgenxtmnnew} is invalid
since it is defined at the singular point $m=0$
of the 4th-order regularized Green's functions.
The occurrence of the singular term and its associated
trace anomaly are artifacts
brought about by the 4th-order subtraction term.
In contrast, the 2nd-order and 0th-order
regularized Green's functions,  \eqref{regGxi0} and  \eqref{G16},
 contain no such kind of singular terms.

\begin{figure}[htb]
\centering
\subcaptionbox{}
    {  \includegraphics[width = .45\linewidth]{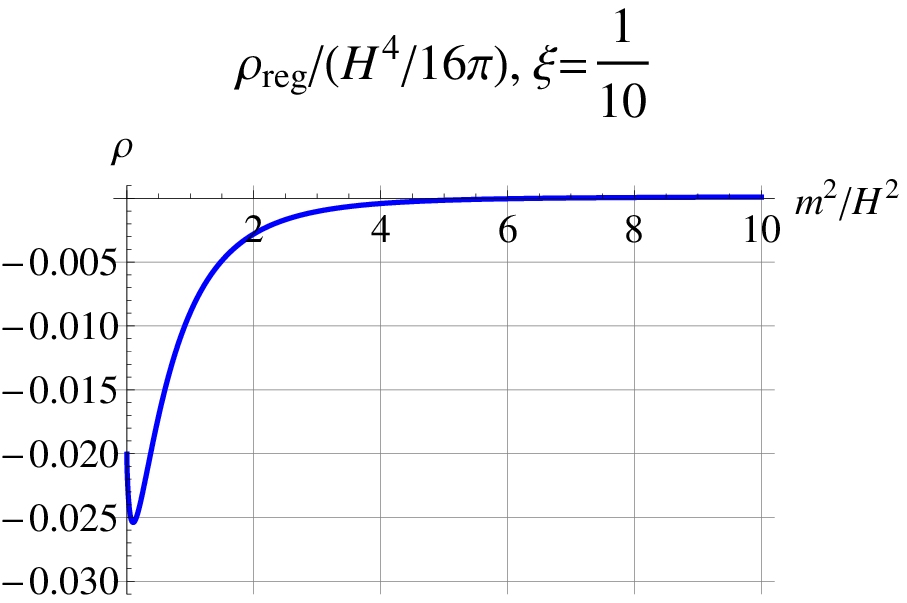}   }
\subcaptionbox{}
    {  \includegraphics[width = .45\linewidth]{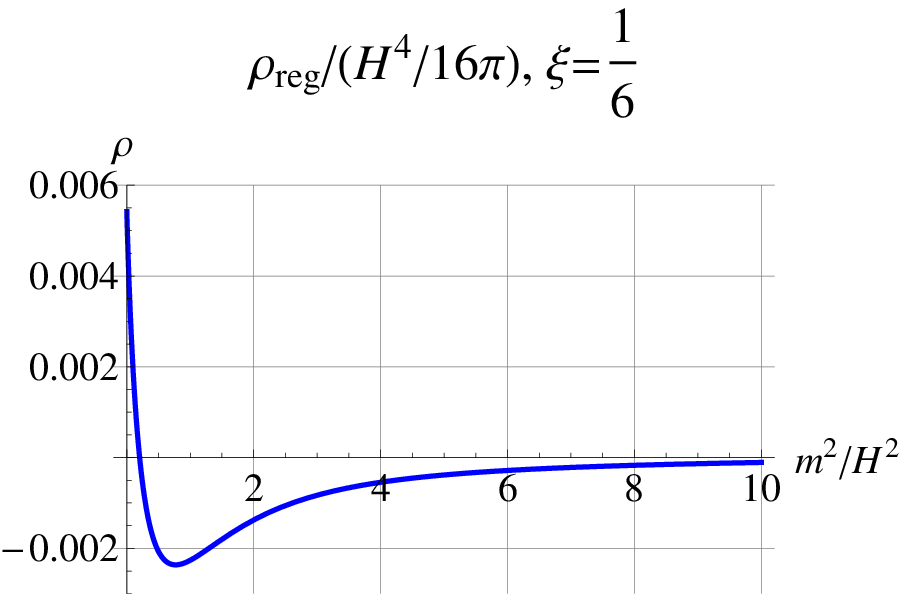}    }
\caption{
(a):  For $\xi=\frac{1}{10}$,
the 4th-order $\rho_{reg}$
is negative in the whole range of $\frac{m^2}{H^2}$.
(b):  For $\xi=\frac16$,
the 4th-order $\rho_{reg}$
  is negative for large mass $\frac{m^2}{H^2} \geq 0.2$.
}
 \label{rho4thxi16}
\end{figure}

Next  the resulting energy density of  \eqref{trted4thxitmunureg}
is generally  negative
\[
\rho_{reg} < 0 ,
\]
as shown  in Fig.\ref{rho4thxi16} (a) for $\xi=\frac{1}{10}$,
  and in Fig.\ref{rho4thxi16} (b) for $\xi=\frac{1}{6}$.
It is checked that  the 4th-order adiabatic
$\rho_{reg} =\int_0^{\infty }(\rho_{k } -\rho_{k~A4}) \frac{d k}{k}$
is also equal to the regularized energy density  of \eqref{trted4thxitmunureg}.
Such a  negative vacuum energy
is inconsistent with  the de Sitter inflation
that requires a positive vacuum energy.
This is another vital difficulty of the 4th-order regularization.

From the above examinations it is clear that
both the trace anomaly and the negative energy density
are simultaneously caused by the over-subtraction of
the 4th-order regularization
which is discordant with the minimum subtraction rule.
Hence, the 4th-order  regularization,
either the adiabatic or the point-splitting,
\cite{DowkerCritchley1976,Brown1977,Hawking1977,Wald1978,BunchDavies1978,
BunchChristensenFulling1978,Bunch1979,Bunch1980}
is an improper prescription for a massive scalar field in de Sitter space.

\section{Conclusion and Discussions}

We have carried out the point-splitting regularization
of the stress tensor of the coupling massive scalar field
in de Sitter inflation.
The key of any regularization
is to prescribe an appropriate subtraction term.
In the point-splitting method,
the stress tensor is constructed from the Green's function in $x$-space,
so the regularized Green's function will be instrumental.
In our previous work  \cite{ZhangYeWang2020},
the 2nd- and 0th-order adiabatically regularized Green's functions
with the coupling $\xi$ were obtained, and are used in this paper.
For a given $\xi$,
assuming the minimal subtraction rule \cite{ParkerFulling1974},
we have performed regularization on
the stress tensor to the same adiabatic order as on the Green's function,
and in two alternative  schemes:
one is to calculate the regularized stress tensor from
the regularized Green's function,
another is to calculate the unregularized,
and subtraction stress tensors respectively
and then to take their difference.
In both schemes,
we have found that,  for $\xi\ne 0$,
the point-splitting calculation may not automatically lead to
an appropriate regularized stress tensor
even when the  regularized Green's function
is continuous and UV- and IR-convergent.
After dropping unwanted higher-order terms,
both schemes yield the same stress tensor,
which is also equal to the outcome from adiabatic regularization.
Comparatively, the second scheme involves more calculations
of the divergent terms,
and, nevertheless, is easier to pick out the unwanted higher-order terms.

For the minimal coupling  $\xi=0$ in Sections 4,
we adopt the 2nd-order regularization
which is sufficient to remove all the UV divergences
and in accordance with the minimum subtraction rule.
The 0-order regularization would not be able to remove all UV divergences,
and the 4-order regularization would subtract off too much and
lead to a negative spectral energy density.
Using the 2nd-order regularized  Green's function,
we have carried out differentiations and the coincidence limit,
and obtained the 2nd-order regularized vacuum stress tensor \eqref{Lambda0},
which is finite and constant,
satisfies the maximal symmetry in de Sitter space,
respects the covariant conservation,
and its energy density is positive.
Thus, it is identified as, or part of,
 the cosmological constant \eqref{Lambda0}.
The special case $m=\xi=0$ needs a separate treatment
in the point-splitting regularization,
and the regularized vacuum Green's function and stress tensor are zero,
the same as the result from the adiabatic regularization.

The case of general  $\xi >0$ in Sections 5
is more involved than the case $\xi=0$.
The coupling $\xi R$ causes a term $\sim \epsilon^2\ln\epsilon^2$ in
the 2nd-order regularized  Green's function,
and consequently brings a divergent term $\sim \ln\epsilon^2$
and other unwanted 4th-order terms  in the regularized stress tensor.
To avoid these, we remove the 4th-order terms
from the subtraction stress tensor,
just as we did in the adiabatic regularization.
There is still a 4th-order term
which depends upon the path of the coincidence limit
and does not possess  the maximum symmetry.
After dropping this path-dependent  term,
the regularized stress tensor \eqref{trace2ndreg2} becomes appropriate,
and reduces to \eqref{Lambda0} when $\xi =0$.
In particular, we have found that,
for small couplings, for instance $\xi  \in(0,\frac{1}{7.04})$
at a fixed $\frac{m^2}{H^2}=0.1$,
the 2nd-order regularized energy density is positive,
and  also can be identified as the cosmological constant,
like the $\xi=0$ case.
But, for large couplings,
say  $\xi > \frac{1}{7.03}$ at a fixed $\frac{m^2}{H^2}=0.1$,
the regularized energy density and spectral energy density
 will still be negative.

For  $\xi=\frac16$ in Sections 6, we adopt the 0th-order regularization
which removes all the UV divergences
and is in accordance with the minimum subtraction rule.
If the 2nd-, or 4th-order regularization were adopted for $\xi=\frac16$,
one would get a negative spectral energy density
and a negative energy density.
By the trace relation \eqref{TGrel16} and by the maximum symmetry,
the  0th-order regularized stress tensor \eqref{mmx} follows
straightforwardly without carrying out differentiations.
Its energy density is positive,
and also can be identified as, or part of,  the cosmological constant.
Alternatively,  we have directly calculated the stress tensor,
and, after extra treatments in analogy to the case  $\xi > 0$,
also arrived at \eqref{mmx}.
In the massless limit,
the regularized vacuum stress tensor is zero,
and  there is no trace anomaly for
the massless scalar field with $\xi=\frac16$.

The conventional 4th-order regularization is also examined in Section 7.
We have  calculated
the 4th-order regularized Green's function and stress tensor with general $\xi$.
We have demonstrated in Fig.\ref{rho4thxi16} that
the 4th-order regularized vacuum energy density for  a general $\xi$
is negative,
which is inconsistent with the de Sitter inflation
that requires a positive vacuum energy.
Moreover,
the 4th-order regularized Green's function \eqref{G04t} is singular at $m=0$,
and consequently its associated  trace anomaly for $\xi=\frac16$
is ill-defined in the massless limit.
These difficulties are caused by the over-subtraction
under the conventional 4th-order regularization
which is discordant with the minimum subtraction rule.

We now discuss the issue of the order of regularization.
The outcome of our paper indicates that
the order of regularization is very important
in achieving an appropriate regularized stress tensor
with the desired properties.
However, there are little discussions
on the issue of order of regularization in literature,
even though there have been
many of studies on regularization since 70's,
and almost all adopted the 4th-order for the stress tensor,
by default, or implicitly.
It seems to us that
there is no unique recipe for the order of regularization
except the desired properties of the stress tensor
that we want to achieve.
In this regard, the most closely related is
the minimum subtraction rule suggested by Ref.\cite{ParkerFulling1974}
that only the minimum number of terms should be subtracted.
The regularization order is actually implied by this rule,
particularly, in the adiabatic regularization method,
by which the subtraction terms are effectively grouped by the orders.
In our paper,
for $\xi=\frac16$, the 0th-order regularization
 yields an appropriate stress tensor
and is in accordance with this rule.
Similarly, for $\xi=0$,
the 2nd-order also works and is also in accordance with this rule.
If the 4th-order were adopted for $\xi=\frac16$ and $\xi=0$,
it is discordant with the minimum subtraction rule,
so as to yield a negative spectral energy density.
Nevertheless, this does not rule out the 4th-order,
which may be necessary in other cases.
Our work has shown that
an appropriate choice of the regularization order
depends upon the coupling.
In general, we speculate that
this may depend also upon the type of quantum fields \cite{ZhangYe2022}
and  the symmetry of spacetime background, etc.
All one can do is by trial and error, in each concrete case.

There  occurs another issue of
the regularization order for the Green's function.
As far as we know, Ref. \cite{Parker2007} first performed
the 2nd-order adiabatic regularization on the power spectrum
for the $\xi=0$ massive scalar field.
In  the point-splitting  regularization,
for the scalar field,
 $\langle T_{\mu\nu} \rangle$ is actually constructed from  $G(x-x')$,
and contains typical terms like $\xi R \, G(x-x')$ and  $m^2 G(x-x')$, etc,
and a regularization of Green's function implies
a regularization of stress tensor.
Therefore, it is natural to conjecture
that the order of regularization
on  the Green's function should be equal to that on the stress tensor.
Indeed, as our calculation shows,
for $\xi=\frac16$ the 0th-order regularization
works for both $\langle T_{\mu\nu} \rangle$ and $G(x-x')$,
and analogously for $\xi=0$ the 2nd-order also works
for both $\langle T_{\mu\nu} \rangle$ and  $G(x-x')$.
This is also true  in the adiabatic regularization
on the scalar field  \cite{ZhangYeWang2020,ZhangWangYe2020}.
In these cases, both methods support
the same  order for the stress tensor and Green's function.
Nevertheless, this conjecture may not hold for
other type of fields, such as vector fields and tensor fields,
for which the Green's functions posses multi components
and the stress tensors is composed of several portions
with different structure \cite{ZhangYe2022}.

Comparing the two methods of regularization,
the point-splitting in this paper and
the adiabatic in Ref.\cite{ZhangYeWang2020,ZhangWangYe2020},
we see the following.

In the adiabatic regularization in $k$-space,
one is able to get the subtraction terms
 to any desired order by the WKB approximation systematically,
for the power spectrum and for the spectral stress tensor.
On the other hand,
in the point-splitting regularization in position space,
the subtraction term for the  Green's function
valid on the whole range
is generally hard to find directly.
The conventional Hadamard function as a subtraction term
is only an approximation at small distance,
and not valid  on the whole range.
With the help  of
the adiabatically regularized power spectrum,
through the Fourier transformation,
one will be able to get the adiabatically regularized
 Green's function.
However, even when the regularized Green's function is given
with the coupling $\xi R \ne 0$,
one still needs extra treatments to drop certain higher-order terms
from the subtraction stress tensor,
and to drop the unwanted path-dependent terms from
the regularized  stress tensor.

In regard to the outcome,
 the two methods are complementary.
The adiabatic regularization yields the regularized spectral stress tensor
and the numerical,  regularized stress tensor after  $k$-integration.
The point-splitting regularization
yields the analytical, regularized stress tensor,
but  not the spectral stress tensor.

\

\textbf{Acknowledgements}

Y. Zhang is supported by NSFC Grant No. 11675165,  11633001,  11961131007,
 and in part by National Key RD Program of China (2021YFC2203100).
B. Wang is supported by
the National Key R\&D Program of China (2021YFC2203100),
NSFC Grants No. 12003029,
the Fundamental Research Funds for the Central Universities
under Grant No. WK2030000044.

\appendix

\section{Some differentiation formulae }

In this appendix, we list some formulae of the point-splitting method
which are used in calculation of the stress tensor in the context.
For simple notation, we introduce
\bl
\epsilon^2 \equiv \frac{\sigma}{2H^2}
 = \frac{(\tau-\tau')^2-(x-x')^2-(y-y')^2-(z-z')^2}{4H^2\tau\tau'} ,
\el
which is the one quarter  of the square of the geodesic  distance
in the de Sitter space, and obeys the equation
$\epsilon^2=(\epsilon^2)_{,\, \mu}(\epsilon^2)^{,\, \mu}$ at small separation.
Performing  differentiations and then taking the coincidence limit,
one  obtains the  basic formulae
\bl
& \lim_{x' \rightarrow x}\nabla_ \mu   \nabla_{\nu } \epsilon^2
= \lim_{x' \rightarrow x}\nabla_ {\mu'}   \nabla_{\nu' } \epsilon^2
= - \lim_{x' \rightarrow x}\nabla_ \mu   \nabla_{\nu \, '} \epsilon^2
= \frac12 g_{\mu\nu},
\label{D10}
\\
& \lim_{x' \rightarrow x}\nabla^\sigma   \nabla_{\sigma } \epsilon^2
=\lim_{x' \rightarrow x}\nabla^{\sigma'}   \nabla_{\sigma'} \epsilon^2
= - \lim_{x' \rightarrow x}\nabla^\sigma   \nabla_{\sigma \, '} \epsilon^2
= 2 .   \label{limitsigma2}
\el
The following formulae are also involved  in the context
{  \allowdisplaybreaks
\bl
\lim_{x' \rightarrow x}\nabla_{\mu}\nabla_{\nu'}\epsilon^{-2}
&=\lim_{x' \rightarrow x}(\frac{1}{\epsilon^4}\cdot \frac12 g_{\mu\nu}+
\frac{2}{\epsilon^6}\cdot\partial_{\nu'}\epsilon^2\cdot
\partial_{\mu}\epsilon^2),\label{164}
\\
\lim_{x' \rightarrow x}\nabla_{\mu}\nabla_{\nu}\epsilon^{-2}
&=\lim_{x' \rightarrow x}(-\frac{1}{\epsilon^4}\cdot \frac12 g_{\mu\nu}+
\frac{2}{\epsilon^6}\cdot\partial_{\nu}\epsilon^2\cdot
\partial_{\mu}\epsilon^2+
\frac{1}{\epsilon^4}\Gamma_{\mu\nu}^{\alpha}\partial_{\alpha}\epsilon^{2}),\label{166}
\\
\lim_{x' \rightarrow x}\nabla_{\sigma}\nabla^{\sigma'}\epsilon^{-2}
&=\lim_{x' \rightarrow x}(\frac{1}{2\epsilon^4}
  \cdot (-1)(1-\frac{\tau'}{\tau})^2+\frac{R}{12 \epsilon^2} ),
\label{1755}
\\
\lim_{x' \rightarrow x}\nabla_{\sigma}\nabla^{\sigma}\epsilon^{-2}
& =  \lim_{x' \rightarrow x}\nabla_{\sigma'}\nabla^{\sigma'}\epsilon^{-2}
=\lim_{x' \rightarrow x }(-\frac16\frac{R}{\epsilon^2}) ,
\label{170}
\\
\lim_{x' \rightarrow x}\nabla_{\mu}\nabla_{\nu'}\ln \epsilon^2
&=\lim_{x' \rightarrow x } (-\frac12g_{\mu\nu}\cdot\frac{1}{\epsilon^2}
-\partial_{\mu}\epsilon^2
\partial_{\nu'}\epsilon^2\cdot\frac{1}{\epsilon^4}),
\label{202}
\\
\lim_{x' \rightarrow x }\nabla_{\mu}\nabla_{\nu}\ln\epsilon^2
&=\lim_{x' \rightarrow x} (\frac12g_{\mu\nu}\cdot\frac{1}{\epsilon^2}
-\partial_{\mu}\epsilon^2
\cdot \partial_{\nu}\epsilon^2\cdot\frac{1}{\epsilon^4}-\frac{1}{\epsilon^2}
\Gamma^{\alpha}_{\mu\nu}\partial_{\alpha}\epsilon^2),\label{271}
\\
\lim_{x' \rightarrow x}\nabla_{\sigma}\nabla^{\sigma'}\ln \epsilon^2
&=\lim_{x' \rightarrow x}(-\frac{1}{\epsilon^2}),\label{210}
\\
\lim_{x' \rightarrow x}\nabla_{\sigma}\nabla^{\sigma}\ln \epsilon^2
& =  \lim_{x' \rightarrow x}\nabla_{\sigma'}\nabla^{\sigma'}\ln \epsilon^2
=\lim_{x' \rightarrow x}\frac{1}4(\frac{4}{\epsilon^2}+R).
  \label{281}
\\
\lim_{x' \rightarrow x}\nabla_{\mu}\nabla_{\nu'}(\epsilon^2\ln\epsilon^{2})
&=
\lim_{x' \rightarrow x }   (\frac{1}{\epsilon^2}\partial_ {\mu'}
  \epsilon^2\cdot\partial_{\nu}\epsilon^2
  -\frac{1}{2}g_{\mu\nu}(\ln\epsilon^2+1)),\label{226}
\\
\lim_{x' \rightarrow x}\nabla_ \mu   \nabla_{\nu}(\epsilon^2\ln\epsilon^2)
&=  \lim_{x' \rightarrow x}  (\frac{1}{\epsilon^2}\partial_ \mu
\epsilon^2\cdot\partial_{\nu}\epsilon^2
 +\frac{1}{2}g_{\mu\nu}(\ln\epsilon^2+1)
 ),
\\
\lim_{x' \rightarrow x}\nabla_ \sigma   \nabla^{\sigma '}
(\epsilon^2\ln\epsilon^2)
&=\lim_{x'\rightarrow x}-(3+2\ln\epsilon^2),\label{boxeslne}
\\
\lim_{x'\rightarrow x}\nabla_{\sigma}\nabla^{\sigma}(\epsilon^2\ln\epsilon^2)
&=   \lim_{x'\rightarrow x}\nabla_{\sigma'}\nabla^{\sigma'}(\epsilon^2\ln\epsilon^2)
=\lim_{x'\rightarrow x} (3+2\ln\epsilon^2) .
\label{289}
\el
}

\section{The term depending on the path of coincidence limit }

In Sections 5 and 6,
  $P_{\mu\nu}$ defined by  \eqref{pathdependent}
 shows up in the stress tensor \eqref{dropr2} and \eqref{reg16ttst2}
when the regularized Green's function
contains $\epsilon^2 \ln\epsilon^2$ for   $\xi\ne 0$.
Different paths of coincidence limit lead to different values of
$\lim_{x^{\,\alpha '} \rightarrow x^{\, \alpha }} P_{\mu\nu} $.
For instance,  consider the $00'$-component of
the first term in \eqref{pathdependent},
\bl
\frac{1}{\epsilon^2}\partial_0\epsilon^2\cdot\partial_{0'}\epsilon^2
&= \frac{1}{\epsilon^2}
 \big(-\frac1\tau\epsilon^2+\frac{\tau-\tau'}{2H^2\tau\tau'}\big)
  \cdot \big(-\frac1{\tau'}\epsilon^2-\frac{\tau-\tau'}{2H^2\tau\tau'}\big)
\nn
\\
&= -\frac{1}{H^2\tau\tau'}\frac{(\tau-\tau')^2}{(\tau-\tau')^2
-(x-x')^2-(y-y')^2-(z-z')^2} .
\label{orderlimit}
\el
For  the path  $\tau' \rightarrow\tau$
  followed by   $\vec{r}' \rightarrow \vec{r}$,
\eqref{orderlimit} gives
\bl
&\lim_{\vec{r}'\rightarrow \vec{r}}\lim_{\tau'\rightarrow \tau}
\frac{1}{\epsilon^2}\partial_0\epsilon^2\cdot\partial_{0'}\epsilon^2
=\lim_{\vec{x}'\rightarrow \vec{x}}
-\frac{1}{H^2\tau^2}\frac{0}{-(x-x')^2-(y-y')^2-(z-z')^2}
  =0,\label{tfxs}
\el
for  the path $\vec{r}' \rightarrow \vec{r}$  followed by  $\tau' \rightarrow\tau$,
\eqref{orderlimit} gives
\bl
&\lim_{\tau'\rightarrow \tau}\lim_{\vec{r}'\rightarrow \vec{r}}
\frac{1}{\epsilon^2}\partial_0\epsilon^2\cdot\partial_{0'}\epsilon^2
=\lim_{\tau'\rightarrow \tau}
-\frac{1}{H^2\tau\tau'}\frac{(\tau-\tau')^2}{(\tau-\tau')^2}
=-\frac{1}{H^2\tau^2} =- a^2(\tau) .\label{xfts}
\el
\eqref{tfxs} and \eqref{xfts}  are not equal.
Similarly, other terms of $P_{\mu\nu}$ also depend on the path.

For  $ P_{\mu\nu} $ as a whole,
detailed calculation shows that,
for the path  $\vec{r'}\rightarrow\vec{r}$ followed by $\tau' \rightarrow\tau$,
\bl
\lim_{\tau'\rightarrow \tau}\lim_{\vec{r}'\rightarrow \vec{r}}
 P_{\mu\nu} =a^2(\tau) \, diag(-1,0,0,0).
 \label{appendix137}
\el
For the path $\tau' \rightarrow\tau$
followed by $ x'\rightarrow x$,
and then irrespectively $ y'\rightarrow y$,  $z'\rightarrow z$,
\bl
\lim_{z'\rightarrow z}
\lim_{y'\rightarrow y}
\lim_{x'\rightarrow x}
\lim_{\tau'\rightarrow \tau}
P_{\mu\nu}  =a^2(\tau)  \, diag (0,1,0,0) .
\label{appendix138}
\el
which is not equal to \eqref{appendix137}.
Other paths will give other values of
$\lim P_{\mu\nu}$
which differ from \eqref{appendix137} \eqref{appendix138}.
Moreover, for any path, $\lim P_{\mu\nu}$
is not proportional to the metric $g_{\mu\nu}$,
and does not respect the  maximum  symmetry in de Sitter space.
Thus, $\lim P_{\mu\nu}$ is dropped from the stress tensor.

\end{document}